\documentclass[aps,pra,twocolumn,superscriptaddress,a4paper,floatfix]{revtex4-1}
\usepackage[latin1]{inputenc}
\usepackage{graphicx}
\usepackage{siunitx}
\usepackage{amsmath,amssymb}
\usepackage{hyperref}
\newcommand \beq{\begin{eqnarray}}
\newcommand \eeq{\end{eqnarray}}

\begin{document}

\title{Bosonic superfluid transport in a quantum point contact}

\author{Shun Uchino}
\affiliation{Weseda Institute for Advanced Study, Waseda University, Shinjuku, Tokyo 169-0051, Japan}

\author{Jean-Philippe Brantut}
\affiliation{Institute of Physics, EPFL, 1015 Lausanne, Switzerland}

%\date{\pdfdate}

\begin{abstract}

We present a microscopic theory of heat and particle transport of an interacting, low temperature Bose-Einstein condensate in a quantum point contact. 
We show that, in contrast to charged, fermionic superconductors, bosonic systems feature tunneling processes of condensate elements, leading to the presence of odd-order harmonics in the AC Josephson current. A crucial role is played by an anomalous tunneling process where condensate elements are coherently converted into phonon excitations, leading to even-order harmonics in the AC currents as well as a DC contribution.
At low bias, we find dissipative components obeying Ohm's law, and bias-independent nondissipative components, in sharp contrast to fermionic superconductors. Analyzing  the DC contribution, we find zero thermopower and Lorenz number at zero temperature, a breakdown of the bosonic Wiedemann-Franz law. These results highlight importance of the anomalous tunneling process inherent to charge-neutral superfluids. The consequences could readily be observed in existing
cold-atom transport setups.

\end{abstract}

\maketitle

\section{Introduction}
A mesoscopic system connected to reservoirs is one of the most important examples of nonequilibrium quantum statistical physics~\cite{nazarov2009}.
For a long time, such a system has mainly been discussed in electron systems, which revealed
nontrivial outcomes absent in bulk systems such as conductance quantization~\cite{datta1997}, current noise~\cite{blanter2000}, 
and fluctuation relation~\cite{RevModPhys.81.1665}.
Recently, atomic quantum gases have emerged as an alternative route to investigate mesoscopic systems~\cite{Chien:2015ab,krinner2017} in particular single-mode quantum point contacts (QPCs)~\cite{krinner2015}.  
QPCs are the corner stone of mesoscopic quantum devices, as building blocks for quantum coherent devices such as quantum dots and interferometers. Their counterparts in atomic gases opens the perspective of complex, quantum coherent 'atom-tronic' devices \cite{Seaman:2007aa,Ramanathan:2011aa,Jendrzejewski:2014aa,Ryu:2015aa} 
featuring controlled interactions~\cite{RevModPhys.80.885} and the possibility to use fermionic or bosonic statistics~\cite{PhysRevB.85.125102,PhysRevA.90.023624,PhysRevA.98.043623}.

In contrast with superconductors which are charged, atomic quantum gases are charge neutral. As a result, in the superfluid phase, there always exist gapless collective modes which are not present in charged systems due to the Anderson-Higgs mechanism~\cite{altland2010,nagaosa2013}. The fact that the dominant low-energy excitations of superconductors are of pair-breaking type has spectacular consequences such as multiple Andreev reflections (MARs) in superconductor-normal interfaces \cite{Zagoskin:1998aa} or QPCs \cite{PhysRevLett.75.1831}, yet little is known concerning the role played by gapless collective excitations for quantum transport in charge-neutral superfluids. In the case of a Bose-Einstein condensate (BEC), relevant to two-terminal systems of cold atoms \cite{PhysRevA.93.063619,krinner2017}, as well as superfluid helium in nano-pores \cite{Duc:2015aa}, a hydrodynamic description correctly predicts the Josephson dynamics \cite{PhysRevLett.95.010402,levy2007,PhysRevLett.106.025302,Xhani:2019aa} and dissipation associated with topological excitations in wide junctions, but fails in the case of a QPC. 

In this paper, we present a  microscopic theory of low-temperature transport of interacting neutral bosons through a single-mode QPC.
Modeling the system with the tunneling Hamiltonian~\cite{PhysRevB.54.7366,PhysRevLett.92.127001,PhysRevB.71.024517,PhysRevLett.118.105303,PhysRevA.98.041601,han2018} and the Bogoliubov theory of interacting condensates~\cite{RevModPhys.76.599,pitaevskii2016}, we use the Keldysh formalism~\cite{rammer2007,kamenev2011} to derive a current formula applicable to heat and particle transport up to a nonlinear response regime.
 %Our theory accounts for interplay between condensate and phonon modes, which is discarded in previous approaches \cite{Gutman:2012aa,PhysRevLett.113.170601,Papoular:2016aa,PhysRevA.98.043623}. 
 We elucidate the crucial role of bosonic enhancement and the gapless Bogoliubov modes for superfluid transport, which is revealed in the differences between odd and even harmonics of the AC Josephson current.  %the crucial contributions 
 %By applying it to particle transport, we show that
 %due to the bosonic enhancement, the dominant transport processes are associated with the condensate and are different between odd and even harmonics in the current. 
 We predict dissipative components obeying Ohm's law even at zero temperature with a fully superfluid system, which is in sharp contrast with the case of charged superconductors~\cite{PhysRevB.54.7366}.
Then, we consider the DC transport of heat and particle currents, which in spite of obeying Ohm's law~\footnote{We note that transport coefficients 
depend on
biases in non-Ohmic cases, where the Lorenz number cannot solely be expressed in terms of fundamental constants.} shows a large breakdown of the bosonic Wiedemann-Franz law. Finally, we propose experiments to confirm our findings in existing two-terminal setups of ultracold atomic gases. 

\begin{figure}[t]
\begin{center}
\includegraphics[keepaspectratio, width=5.5cm,clip]{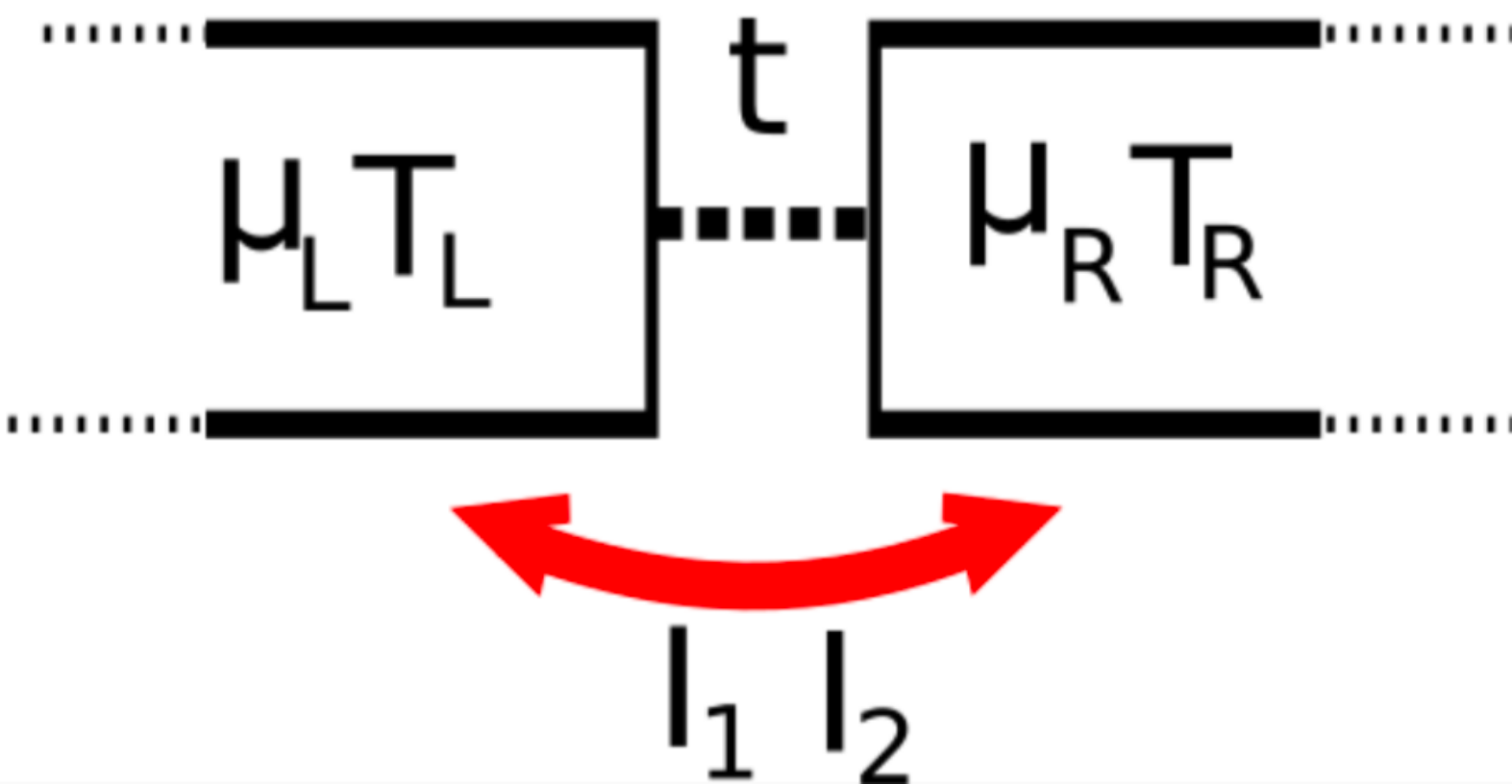}
\caption{Model used for the calculation, with $T_i$, $\mu_i$ the temperatures and chemical potential in reservoir $i$, $t$ the transmission amplitude for single atoms across the contact and particle and heat currents $I_1$ and $I_2$ respectively. 
\label{fig:setup}} 
\end{center}
\end{figure}

This paper is organized as follows. 
In Sec. II, we present the model and methods yielding an expression for the mesoscopic current, which is the basis of this work.
In Sec. III we use this expression to obtain results for the AC and DC components of the current at fixed biases.
In Sec. IV,  we present predictions for cold-atom experiments, including the effect of the trapping potential, for the time evolution of thermodynamic quantities and Shapiro resonances.
Section V summarize results and mentions possible applications of this work. In appendices, some technical details on the theoretical analyses are provided.

\section{Model and methods}
We consider a two-terminal transport system, with two macroscopic reservoirs~\cite{PhysRevLett.110.200406} filled by a weakly-interacting BEC and connected by a mesoscopic channel, as shown in figure \ref{fig:setup}. We describe the channel as a single mode QPC, which is adequate when its width and lengths are much shorter than the coherence length of the BEC. In such a case, the details of the motion of atoms in the constriction become irrelevant~\cite{RevModPhys.51.101,PhysRevLett.75.1831}, which allows one to start with the following tunneling Hamiltonian~\cite{PhysRevB.54.7366,mahan2013,PhysRevLett.92.127001,PhysRevB.71.024517,PhysRevB.84.155414,PhysRevLett.118.105303,PhysRevA.98.041601,han2018}:
\begin{eqnarray}
H=K_L+K_R+t[\psi^{\dagger}_L(\mathbf{0})\psi_R(\mathbf{0})+\psi^{\dagger}_R(\mathbf{0})\psi_L(\mathbf{0})],
\label{eq:hamilton}
\end{eqnarray} 
where   $\psi_{L(R)}(\mathbf{x})$ and $K_{L(R)}$ are the  field operator of the reservoir $L (R)$ and the
grand-canonical Hamiltonian of a Bose gas 
($K_{L(R)}=H_{L(R)} -\mu_{L(R)}N_{L(R)}$ with the Hamiltonian operator $H_{L(R)}$ and with the number operator $N_{L(R)}$), respectively. The last two terms in Eq.~\eqref{eq:hamilton} describe exchanges of particles between the reservoirs, where $t$ is the tunneling amplitude.
In the two-terminal system, the particle and heat current operators can respectively be expressed as $I_1=-\dot{N}_L$ and $I_2=-\dot{K}_L$. 

We consider a regime of weak interaction and high degeneracy, and describe the BECs in the reservoirs with the Bogoliubov theory~\cite{RevModPhys.76.599,pitaevskii2016}.  
To construct a theory of heat and particle transport, we adopt the Keldysh formalism~\cite{rammer2007,kamenev2011}, allowing for a description of the dynamics in the presence of arbitrary bias and tunneling amplitude.
Under fixed chemical potential bias ($\Delta\mu$) and temperature bias ($\Delta T$), the particle and heat currents at time $\tau$ can be expanded 
as a Fourier series  in harmonics of $\Delta\mu$ and be obtained as (see Appendix A for derivation),
\begin{eqnarray}
I_q(\tau)&&=\text{Re}\big[\sum_{m\in \mathbb{Z}}I_{q,2m-1}e^{(2m-1)i\phi(\tau)}+\sum_{m\in \mathbb{Z}}I_{q,2m}e^{2mi\phi(\tau)}\big] \nonumber\\
&&\equiv I_q^{\text{odd}}(\tau)+I_q^{\text{even}}(\tau), \label{eq:current}\\
I_q^{\text{odd}}&&=2\text{Re}[(\{i\partial_{\tau}-\Delta\mu\}^{q-1}\hat{T}^R_{RL})\circ\hat{g}^<_{LR}\circ\hat{T}^A_{RL}\hat{t}^{-1}\nonumber\\
&&\ \ +i\hat{t}(\partial^{q-1}_{\tau} \hat{g}^R_{LL})\circ\hat{T}^R_{LR}\circ\hat{g}^<_{RL}\circ\hat{T}^A_{LR}\circ\hat{g}^A_{RR}]_{11},\label{eq:odd-current}\\
I_q^{\text{even}}&&=2\text{Re}[i
(\partial^{q-1}_{\tau}\hat{g}^R_{LL})\circ\hat{T}^R_{LR}\circ\hat{g}^<_{RR}\circ\hat{T}^A_{RL} \nonumber\\
&&\ \ +(\{i\partial_{\tau}-\Delta\mu\}^{q-1}\hat{T}^R_{RL})\circ\hat{g}^<_{LL}\circ\hat{T}^A_{LR}\circ\hat{g}^A_{RR} ]_{11}, \label{eq:even-current}
\end{eqnarray}
where $q=1$ or 2, $\partial_{\tau}\phi(\tau)=\Delta\mu$ and
$\circ$ represents integration over the internal time variable from minus infinity to plus infinity.
We note that in this paper we set $\hbar=k_B=1$ except for plots of figures.
In addition, $\hat{g}^R$,  $\hat{g}^A$,  and  $\hat{g}^<$ are uncoupled retarded, advanced, and lesser Green's functions, which have $2\times2$ structures
 conventionally employed in a BEC system~\cite{rammer2007,fetter2012}. 
 The Keldysh formalism incorporates the coherent sum of all
 processes by which atoms can traverse the channel. 
 Such a beyond-linear-response effect is accumulated in the renormalized hopping matrix $\hat{T}^{R(A)}$ defined as
 \begin{eqnarray}
 \hat{T}_{RL}^{R(A)}=\hat{t}+\hat{t}\circ\hat{g}^{R(A)}_{LL}\circ
 \hat{t}^{\dagger}\circ\hat{g}^{R(A)}_{RR}\circ \hat{T}_{RL}^{R(A)},
\label{eq:tunnel}
 \end{eqnarray}
 where $\hat{T}^{R(A)}_{LR}(\tau,\tau')=[\hat{T}^{A(R)}_{RL}(\tau',\tau) ]^{\dagger}$ and
 $\hat{t}(\tau,\tau')=\text{diag}(te^{-i\phi(\tau)},te^{i\phi(\tau)} )\delta(\tau-\tau') $.
 Although the structure is similar to that of superconducting QPCs, in that the reservoir Green's functions are $2\times 2$ matrices, the emergence of $I^{\text{odd}}_q$ is peculiar to the bosonic system. In fermionic systems $\hat{g}^<_{LR(RL)}=\hat{0}$ and odd harmonics are absent, because fermionic superfluidity involves pairing. 
The current expression obtained above is the basis for the analysis of mesoscopic BEC systems. In the weak coupling regime, it describes in particular a BEC in a double-well~\cite{PhysRevLett.95.010402,levy2007,PhysRevLett.106.025302} as long as motion along finite extension of the junction can be neglected.

 \begin{figure}[t]
\begin{center}
\includegraphics[width=8.5cm]{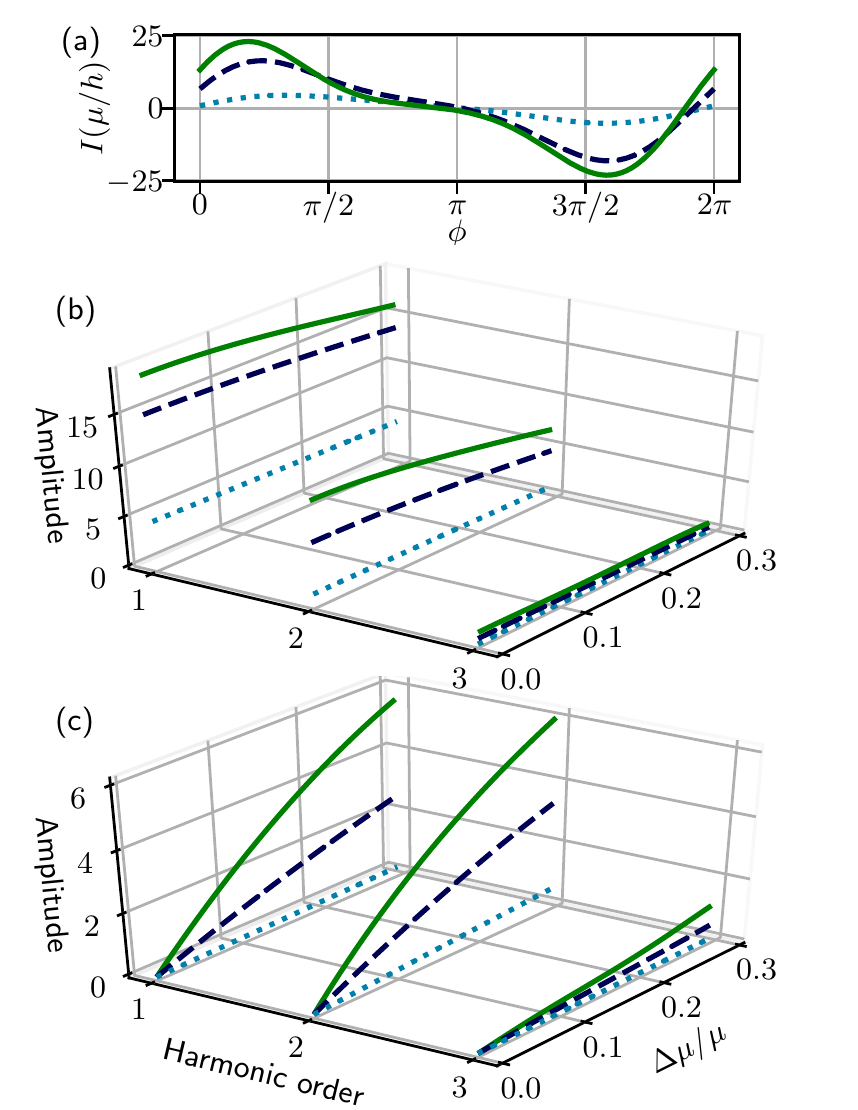}
\caption{(a) AC component of the particle current $I_1(\phi(\tau))=\sum_{m\ne0}[I^{\text{d}}_{1,m}\cos(m\phi)+ I^{\text{nd}}_{1,m}\sin(m\phi)]$ for different tunneling amplitudes $t\rho(\mu)=0.5$ (solid curve), $0.4$ (dashed curve), and 0.2(dotted curve) for $\Delta \mu/\mu = 0.3$. The current is plotted in units of $\mu/h$.
(b) The first three components of the nondissipative current amplitude $I^{\text{nd}}_{1,m}$ as a function of $\Delta\mu$. 
(c) The first three components of the dissipative current amplitude $I^{\text{d}}_{1,m}$.
\label{fig:AC}}
\end{center}
\end{figure}

\begin{figure}[t]
\begin{center}
\includegraphics[width=9cm]{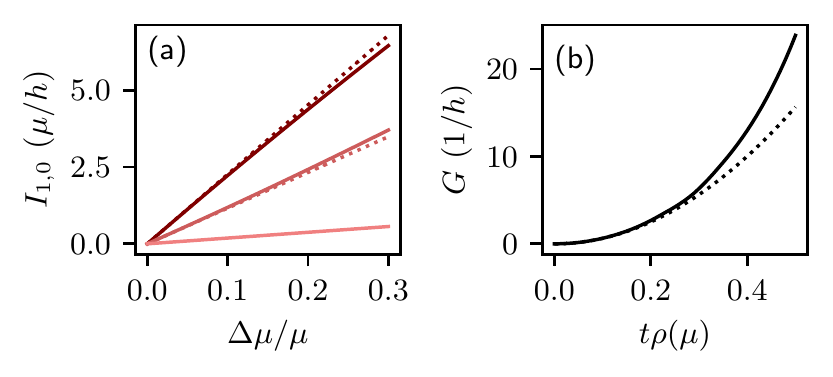}
\caption{(a) DC current $I_{1,0}$ as a function of the bias normalized by $\mu$ for different tunneling amplitudes $t\rho(\mu)=0.5$ (dark red), $0.4$ (red) and $0.2$ (light red). The solid lines shows the numerical solution, the dashed lines show a linear fit to the data. For $t\rho(\mu)=0.2$ it is not distinguishable from the exact result. (b) Conductance as function of tunneling. The solid line shows the exact result, the quadratic dashed line is the result of the linear response theory. 
\label{fig:linearIV}}
\end{center}
\end{figure}

\section{Results}
In this section %in order to reveal current-bias characteristics of a bosonic superfluid quantum point contact,
we consider the situation in which the chemical potential in the reservoirs is fixed, which is standard in the condensed-matter physics context.
In order to connect to quantum gas experiments, in what follows we present figures for which an average chemical potential $\mu=400$nK, the scattering length $a=25$nm, and average density in reservoirs $n=10^{19}$m$^{-3}$.%, are used, consistent with typical numbers for experiments using molecular BECs of $^6$Li.

\subsection{AC component}
A first outcome of our formalism is the structure of the AC components of the particle current at zero temperature.
Due to bosonic enhancement, the dominant transport process is associated with the condensate elements.
In the Bogoliubov prescription~\cite{fetter2012},  the field operator is decomposed into
the the c-number part describing the condensate and the rest describing the non-condensate elements,
and so is Green's function. In our formulation, such a c-number part (condensate element) appears in lesser Green's function
and is expressed as
 \begin{eqnarray}
 \hat{g}^<_{\alpha \beta}(\tau)\to-\frac{ i\sqrt{\mu_{\alpha}\mu_{\beta}}}{g}
\begin{pmatrix} 1 & 1 \\1 & 1
\end{pmatrix},
\label{eq:sub}
\end{eqnarray}
where $\alpha,\beta=$L or R, and $g$ is the coupling constant of the interatomic interaction~\footnote{The c-number part  of the field operator 
does not evolve in time and so is that of lesser Green's function~\cite{fetter2012}.}.
We note that in contrast to the phonon contribution of lesser Green's function shown in Appendix A,
Eq.~\eqref{eq:sub} does not depend on the Bose distribution function being a function of temperature.  
Thus, the processes associated with the condensate survive at zero temperature.

In the tunneling limit where $\hat{T}_{RL}\to\hat{t}$, the leading current term is linear in $t$. It arises from the $I_{1,-1}$ component, and the corresponding current is $I_1(\tau)\to-\frac{2\sqrt{\mu_L\mu_R}t}{g}\sin\phi(\tau)$.
This is the Josephson current associated with the tunneling of condensate elements between reservoirs and is
consistent with the result in a double-well BEC~\cite{PhysRevLett.95.010402,PhysRevLett.106.025302,pitaevskii2016}.

The replacement $\hat{T}_{RL}\to\hat{t}$ also allows to discuss the current component quadratic in $t$. This corresponds to the linear response analysis~\cite{PhysRevA.64.033610}, and yields a DC component $I_{1,0}$ and AC components $I_{1,\pm2}$. An analytic calculation discussed in Appendix B shows the presence of dissipative and non-dissipative terms proportional to $\cos2\phi(\tau)$, $\sin2\phi(\tau)$ respectively. The dominant microscopic process in even harmonics is the tunneling of a condensate element coherently converted into phonon excitation. This anomalous tunneling process is fundamentally different from the tunneling process of the non-dissipative odd harmonics and from tunneling processes in the superconductor case. 

To include systematically higher order tunneling effects with all the virtual processes, we numerically solve Eqs.~\eqref{eq:current}-\eqref{eq:tunnel} under the low-energy expression of $\hat{g}^{R(A)}$ shown in Appendix A.
Figure~\ref{fig:AC} depicts the results of first three AC components for various tunneling amplitudes normalized by the density of states in the reservoirs at the chemical potential $\rho(\mu)$, as in the case of fermions~\cite{PhysRevB.54.7366}.
As can be seen from Fig.~\ref{fig:AC}(a), the AC current deviates from sinusoidal behavior with increasing $t$, revealing the important role of higher harmonics.

Moreover, Fig.~\ref{fig:AC}(c) demonstrates the presence of the dissipative term proportional to $\cos \phi(\tau)$, which is absent at the linear response level. Remarkably, our calculation shows that the amplitudes in the dissipative components obey Ohm's law at small $\Delta\mu$, in contrast to the fermionic case where the nonlinear bias dependence appears~\cite{PhysRevB.54.7366}. This difference originates from $\hat{g}^{R(A)}(\omega)$, which is non-singular for bosons but is singular at the pairing gap frequency for fermions~\cite{PhysRevB.54.7366}. In contrast, we find that the non-dissipative components $\propto\sin m\phi(\tau)$ has a nonzero amplitude even at $\Delta\mu=0$ (Fig.~\ref{fig:AC}(c)), comparable with the fermionic case up to the nonlinearity~\cite{PhysRevB.54.7366}.

\subsection{DC component}
We now focus on the DC component of the particle current $I_{1,0}$ as a function of $\Delta\mu$ at zero temperature. This is directly measurable in cold-atom experiments where all the AC components are averaged out~\cite{husmann2015}. The resulting current-bias relations for different $t$ are shown in figure~\ref{fig:linearIV}a. In all cases, Ohm's law is obeyed in the low-bias regime. This again contrasts with the case of superconductors showing nonlinearities associated with MARs \cite{PhysRevLett.75.1831}. This also differs from the case with one-dimensional reservoirs described by the Luttinger liquid, where power-law current-bias relations are obtained due to critical fluctuations~\cite{Giamarchi:2004aa}.  

Our formalism allows for a systematic investigation of the conductance in the linear regime as a function of the tunneling amplitude. The result is shown as solid line in figure \ref{fig:linearIV}b. Remarkably, even for moderate tunneling amplitudes of the order of $0.4/\rho(\mu)$, the conductance is ten times larger than that of the upper bound of non-interacting fermions~\cite{Beenakker}. In the tunneling regime, conductance shows a quadratic dependence on $t$ (dashed line in figure \ref{fig:linearIV}b), suggesting that the linear response theory accurately captures the physics.

To clarify the nature of the DC transport we now include the heat current, focusing again on the linear response regime.
An analytic calculation  derived in Appendix B shows that
the DC particle and heat currents obey an Onsager matrix,
\begin{eqnarray}
\begin{pmatrix}
I_{1,0}\\
I_{2,0}
\end{pmatrix}
=
\begin{pmatrix}
L_{11} & L_{12} \\
TL_{12} & L_{22}
\end{pmatrix}
\begin{pmatrix}
\Delta \mu\\
\Delta T
\end{pmatrix}.
\label{eq:onsager}
\end{eqnarray}
Here, transport coefficients ($L_{11}, L_{12}, L_{22}$) depend on temperature $T$. They are proportional to $t^2$, by hypothesis. The conductance is expressed as $G=L_{11} = \frac{t^2  \mu^2}{\pi c^3}\left[ \frac{1}{g} + \frac{T^2}{12c^3} \right]$ with the speed of sound $c$. It features a zero-temperature contribution of the condensate inversely proportional to the interaction strength, due to the gapless modes in the reservoirs, and a contribution quadratic in temperature which originates from the incoherent tunneling of Bogoliubov quasiparticles. Importantly, the zero temperature part dominates over the thermal component up to $T \sim  \mu/(na^3)^{1/4} \gg \mu$, where $\sqrt{na^3}$ is the small parameter in the Bogoliubov theory. 

\begin{figure}[t]
\begin{center}
\includegraphics[width=9cm]{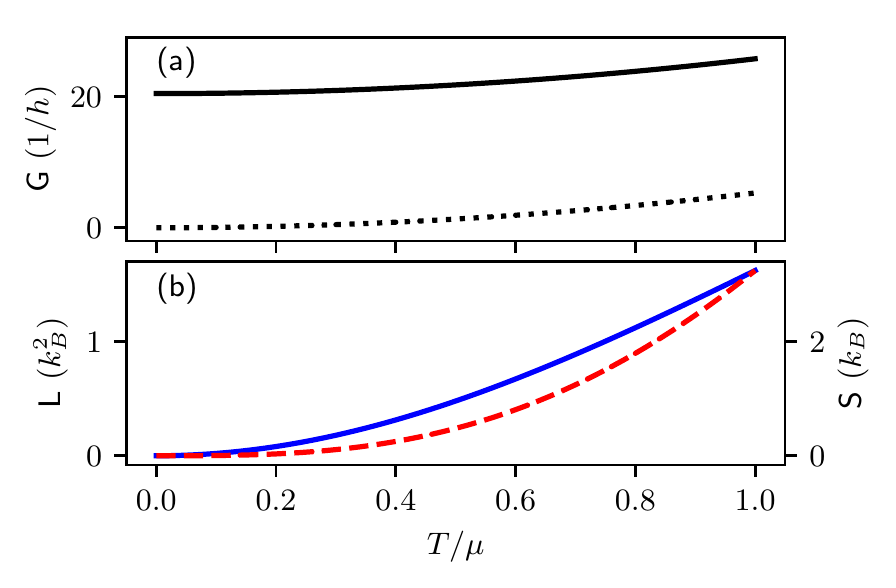}
\caption{Temperature dependence of the transport coefficients (see text for the parameters). (a) Conductance including the phonon contributions (solid lines) and excluding them (dashed lines). The latter is consistent with \cite{PhysRevLett.113.170601}. (b) Lorenz number (blue solid line, left axis) and Seebeck coefficient (red dotted line, right axis), in the tunneling approximation. %The Lorenz number remains significantly below the Wiedemann-Franz law $L=4\pi^2/5\,k_B\sim7.9\,k_B$.
\label{fig:GLS}}
\end{center}
\end{figure}

We also find no contribution from the condensate to heat transport. As a result, the Lorenz number $L=\frac{L_{22}}{TL_{11}}-\frac{L_{12}L_{21}}{TL_{11}^2}$ and Seebeck coefficient $S=\frac{L_{12}}{L_{11}}$, which characterize the relation between entropy and particle currents, go to zero as temperature is reduced. The leading temperature dependence turns out to respectively be quadratic and cubic for the Lorenz number and Seebeck coefficient, again a consequence of the phonon Green's function structure (see Appendix B).  
We note that this behavior of $L$ is quite different from the Wiedemann-Franz law where $L$ is given by a universal number.
Normally, this law is obeyed when charge and entropy are carried by the same fundamental processes such as quasiparticles in Fermi liquids~\cite{ashcroft,Beenakker} and magnons in spin systems~\cite{PhysRevB.92.134425,PhysRevB.98.094430}. 
Even in a BEC at $T\approx T_c$, it is predicted due to dominant quasiparticle transport \cite{PhysRevLett.113.170601}.
Remarkably, we find that even though mass and heat transport originates from quasiparticles, the presence of the anomalous tunneling process yields a large violation of the Widemann-Franz law.

Figure \ref{fig:GLS} illustrates these results, where the conductance (on panel a), Seebeck coefficient and Lorenz numbers (on panel b) are presented as a function of $T/\mu$. Over the whole range of temperatures, the results are dominated by the zero-temperature contribution to current originating from the phonon contribution. Note that the Bogoliubov theory fails around the critical temperature of a BEC, so we cannot extrapolate our results to the high temperature regime covered in particular in \cite{PhysRevLett.113.170601}.

\section{Experimental considerations}
In this section, we discuss the situation of finite reservoirs yielding a finite charging energy, which is
relevant to cold-atom experiments.

\subsection{Time evolution of thermodynamic quantities}
Quantum gas two-terminal or double-well setups, for which our model directly applies, feature finite-sized reservoirs with a chemical potential self-consistently determined from the atom number. As a result, the coupling between the reservoirs competes with their finite compressibility, yielding the crossover between Josephson oscillations and non-linear self-trapping \cite{PhysRevLett.79.4950}. The presence of a finite DC conductance implies directly that the self-trapping regime is not stable, as a system initialized at large bias will slowly relax into the low bias, Josephson oscillation regime. Such a relaxation has been reported in~\cite{PhysRevLett.120.025302,PhysRevLett.123.260402}. 

We now consider a similar scenario for single-channel point-contact transport accounting explicitly for the effects of finite-sized reservoirs. 
%Since the ETH group has already realized a single-mode QPC~\cite{krinner2017}, one can  reach the strong coupling regime (Bose-Einstein condensation regime) by using the Feshbach resonance.
As mentioned before, we fix the parameters ($\mu=400$nK, $a=25$nm, and $n=10^{19}$m$^{-3}$). 
For such a parameter choice with $t\rho(\mu)\sim 0.1$, the ratio of the condensation energy to the Josephson coupling energy turns out to be of the order of $10^6$. 
For a typical initial bias of  
$\Delta N/N\sim0.1$~\footnote{In this case, the transition between Josephson oscillation and self-trapping regimes occurs when
the ratio of the condensation energy to the Josephson coupling energy becomes $10-10^2$.}, expressed in terms of particle number relative imbalance, the system is initially deeply in the self-trapping regime~\cite{PhysRevLett.79.4950,pitaevskii2016}.

We start with a description of the thermodynamics of the reservoirs at equilibrium. 
The differences of thermodynamic quantities between the reservoirs obey the following Maxwell relations:
\beq
\begin{pmatrix}
\Delta N \\
\Delta S
\end{pmatrix}=
\begin{pmatrix}
\kappa & \alpha \\
\alpha & \frac{C_{\mu}}{T}
\end{pmatrix}
\begin{pmatrix}
\Delta \mu \\
\Delta T
\end{pmatrix},
\label{eq:maxwell}
\eeq
where $\Delta N$, $\Delta S$, $\kappa$, $\alpha$, and $C_{\mu}$ are relative atom-number difference, relative entropy difference,
compressibility, dilatation coefficient, and heat capacity at constant $\mu$, respectively.
We note that the thermodynamic coefficients ($\kappa$, $\alpha$, and $C_{\mu}$) are sensitive to geometry of reservoirs.
In a box-type reservoir, they are calculated with the Bogoliubov theory as
\beq
&&\kappa_b\approx\frac{\Omega}{g}+\frac{\pi^2\Omega T^4}{24c^3\mu^{2}},\\
&&\alpha_b\approx-\frac{\pi^2\Omega T^3}{15c^3\mu},\\
&&C_{\mu,b}\approx\frac{2\pi^2\Omega T^3}{15c^3},
\eeq
where $\Omega$ is the volume in each reservoir.
 In contrast,
in a harmonically trapped reservoir, we obtain
\beq
&&\kappa_h\approx\frac{\Omega}{g}-\frac{0.36\Omega T^{7/2}}{\pi c^3\mu^{3/2}},\\
&&\alpha_h\approx\frac{2.5\Omega T^{5/2}}{\pi c^3\mu^{1/2}},\\
&&C_{\mu,h}\approx\frac{13\Omega\mu^{1/2} T^{5/2}}{\pi c^3},
\eeq
where we assume that the trap is isotropic, in which case $\Omega=\frac{4\pi}{3} \Big(\frac{2\mu}{M\omega_{\text{ho}}^2}\Big)^{3/2}$
with the trapping frequency $\omega_{\text{ho}}$.
The temperature and chemical potential responses to variations of internal energy and particle number depend on the shape of the reservoirs, and are thus different for box and harmonic traps. As we show below, this causes different behaviors of the time evolutions of thermodynamic quantities. 

We now describe the slow time evolution of the thermodynamic quantities characterizing the reservoirs resulting from quasi-steady state currents. The quasi-steady approximation, well verified in the experimentally relevant situations where DC transport is concerned, implies that the reservoirs are at thermal equilibrium at each point in time \cite{krinner2017}. This allows for a direct generalization of the linear response results reported above to the case of slow time evolution of biases. 

%These considerations are now combined with the current expressions up to the linear response obtained above, provided that chemical potential and temperature are fixed in time consistent with the quasi-steady state hypothesis.
To deal with the case where biases between the reservoirs can evolve slowly in time,
we adopt the model in which $\phi(\tau)$ is an independent dynamical
variable~\cite{PhysRevA.93.063619,PhysRevLett.120.025302}.
There, in addition to Eq.~\eqref{eq:maxwell}, we consider the following relations:
\beq
&&\frac{1}{2}\frac{d\Delta N}{d\tau}=-I_1, \label{eq:c1}\\
&&\frac{T}{2}\frac{d\Delta S}{d\tau}=-I_2, \\
&&\frac{d\phi}{d\tau}=\Delta\mu.\label{eq:phase}
\eeq
which represent Kirchhoff's law, applicable in the low-frequency regime consistent with the quasi-steady state hypothesis. 
By numerically solving Eqs.~\eqref{eq:maxwell}, and \eqref{eq:c1}-\eqref{eq:phase}, the time evolution of the thermodynamic quantities can be determined.

To discuss the global structure of the time evolution, we also introduce an approximation that neglects the AC components of the currents. 
 In this case, by combining Eq.~\eqref{eq:onsager} and Eq.~\eqref{eq:maxwell}, we obtain
 \beq
 \tau_0\frac{d}{d\tau}
\begin{pmatrix}
\Delta N/\kappa\\
\Delta T
\end{pmatrix}
=-
\begin{pmatrix}
1 & -S_{\text{eff}} \\
-\frac{S_{\text{eff}}}{l} & \frac{L+S_{\text{eff}}^2}{l}
\end{pmatrix}
\begin{pmatrix}
\Delta N/\kappa\\
\Delta T
\end{pmatrix},
\label{eq:linear-eq}
 \eeq
 where $\tau_0=\kappa/(2L_{11})$ is the transport time for mass transport, 
 $l\equiv C_{\mu}/(\kappa T)-(\alpha/\kappa)^2$ is the analog of the Lorenz number for the reservoirs,
and $S_{\text{eff}}\equiv \alpha/\kappa-S$ is the effective Seebeck coefficient. 
Since the equation above is the  first-order differential equation,  the time evolutions can analytically be determined.
The general solution is expressed as
\begin{widetext}
\beq
&&\Delta N(\tau)=\frac{1}{2}\Big\{e^{-\tau/\tau_-}+e^{-\tau/\tau_+} -\Big(1-\frac{L+S^2_{\text{eff}}}{l}\Big)\frac{e^{-\tau/\tau_-}-e^{-\tau/\tau_+}}{\lambda_+-\lambda_-}\Big\}
\Delta N(0)
+\frac{S_{\text{eff}}\kappa(e^{-\tau/\tau_-}-e^{-\tau/\tau_+})}{\lambda_+-\lambda_-}\Delta T(0),\label{eq:general1}\\
&&\Delta T(\tau)=\frac{1}{2}\Big\{e^{-\tau/\tau_-}+e^{-\tau/\tau_+} -\Big(1-\frac{L+S^2_{\text{eff}}}{l}\Big)\frac{e^{-\tau/\tau_-}-e^{-\tau/\tau_+}}{\lambda_+-\lambda_-}\Big\}
\Delta T(0)
+\frac{S_{\text{eff}}(e^{-\tau/\tau_-}-e^{-\tau/\tau_+})}{l\kappa(\lambda_+-\lambda_-)}\Delta N(0).\label{eq:general2}
\eeq
\end{widetext}
Here, the decay-time parameters $\tau_{\pm}$ are expressed as 
\beq
\tau_{\pm}=\tau_0/\lambda_{\pm},
\eeq
with
\beq
\lambda_{\pm}=\frac{1}{2}\Big(1+\frac{L+S^2_{\text{eff}}}{l}\Big)\pm\sqrt{\frac{S^2_{\text{eff}}}{l}+\frac{1}{4}\Big(1-\frac{L+S^2_{\text{eff}}}{l}\Big)^2}.
\eeq

\begin{figure}[t]
\begin{center}
\includegraphics[width=8.cm]{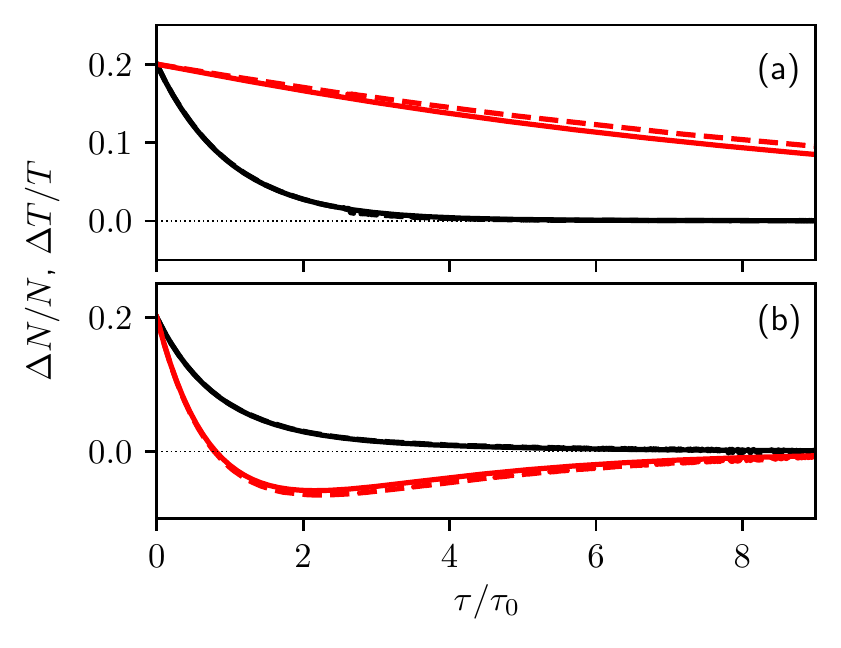}
\caption{Time evolutions of $\Delta N(\tau)/N$ (black) and $\Delta T(\tau)/T$ (red), for initial conditions $\Delta N(0)/N=0.2$ and $\Delta T(0)/T = 0.2$. 
The reservoirs are (a) harmonic traps and (b) box traps. The solid lines describe the full numerical solutions and the dashed lines are the results of Eqs.~\eqref{eq:general1} and~\eqref{eq:general2}.
\label{fig5}} 
\end{center}
\end{figure}

%Based on Eqs.~\eqref{eq:general1} and~\eqref{eq:general2} as well as the numerical calculations including the AC currents,
%we now demonstrate time evolutions of $\Delta N(\tau)$ and $\Delta T(\tau)$.
In Figure~\ref{fig5} we present this result for initial conditions $\Delta N(0)/N=0.2$ and $\Delta T(0)/T=0.2$, together with a numerical solution including the AC components. The parameters mentioned before are used with $T=150$nK and $t\rho(\mu)=0.2$.
Under these parameters, regardless of trapped geometries, we obtain $\tau_0\sim10$s similar to the case of a single mode conductor for fermionic atoms~\cite{krinner2017}. The large conductance of the BEC compared with one expected for noninteracting fermions is compensated by the much larger compressibility of the reservoirs, yielding similar timescales. This illustrates the important role of the finite-size reservoirs in the interpretation of the experiments.

In the case of harmonically trapped reservoirs, the initial relative atom number difference decays in the time scale $\tau_0$ as shown
in Fig.~\ref{fig5}(a).
However, the low values of Lorenz number and Seebeck coefficient at the low temperature, and the near cancellation of the effective Seebeck coefficient between the QPC and reservoir contributions~\cite{Grenier:aa}
lead to the slower decay of the initial temperature difference.
In the case of box reservoirs~\cite{PhysRevLett.110.200406,Mukherjee:2017aa,Luick:2019aa}, 
on the other hand, both of $\Delta N$ and $\Delta T$ decays in the time scale $\tau_0$
despite the low values of Lorenz number and Seebeck coefficient.
The faster decay of $\Delta T$ in box reservoirs is due to the fact that  $L/l\approx0.5$, that is,
the small Lorenz number is compensated by the low value of the analog of Lorenz number for reservoirs.
Finally, we note that in large times where $\tau/\tau_0\gg1$ and $\Delta N/N\approx \Delta T/T\approx0$, the effect of the AC oscillations can be seen in
the numerical calculations. Except for this effect, the
overall differences between analytic solutions (dashed curve) and numerical solutions with the AC components (solid curve) are minuscule.
%This suggests that the analyses without the AC components capture the macroscopic features of the decay dynamics of thermodynamic quantities.

\subsection{Shapiro resonance as a probe of AC components}
The effects of the phonon modes on the AC currents could be observed by direct Fourier analysis of the reservoirs population dynamics. This has been demonstrated at low frequency for wide junctions~\cite{levy2007,PhysRevLett.106.025302}, but might reveal difficult for weak harmonics at high frequencies. Therefore, we  propose to circumvent this difficulty by leveraging the time-dependent control over trapping potentials and observe Shapiro resonances~\cite{barone1982,tinkham2004}, mapping the harmonics of the Josephson current onto the DC component. 

To this end, we  consider the time-dependent hopping amplitude such that the
phase parameter is given by
\beq
\phi(\tau)=\phi_0+\Delta\mu\tau+\alpha\sin(\omega_1\tau),
\eeq
where $\omega_1$ and $\alpha$ are the oscillation frequency and its amplitude, respectively.
By substituting the above into the particle current expression, we obtain
\begin{widetext}
\beq
I_{1}(\tau)&&=\sum_{m=0}^{\infty}[I^{\text{nd}}_{m}\sin(m\phi(\tau))+I^{\text{d}}_{m}\cos(m\phi(\tau)) ]\nonumber\\
&&=\sum_{m=0}^{\infty}I^{\text{nd}}_{m}\Big[\sin(m(\phi_0+\Delta\mu\tau))\cos(m\alpha\sin(\omega_1\tau))+\cos(m(\phi_0+\Delta\mu\tau))\sin(m\alpha\sin(\omega_1\tau)\Big]
\nonumber\\
&&+\sum_{m=0}^{\infty}I^{\text{d}}_{m}\Big[\cos(m(\phi_0+\Delta\mu\tau))\cos(m\alpha\sin(\omega_1\tau))-\sin(m(\phi_0+\Delta\mu\tau))\sin(m\alpha\sin(\omega_1\tau)\Big],
\eeq

%By using the following mathematical formulae,
%\beq
%&&\cos(x\sin\theta)=J_0(x)+2\sum_{n=1}^{\infty}J_{2n}(x)\cos(2n\theta),\\
%&&\sin(x\sin\theta)=2\sum_{n=1}^{\infty}J_{2n-1}(x)\sin((2n-1)\theta),
%\eeq
which can be expressed in terms of the Bessel functions of order $n$ $J_n$, as
\beq
I_1(\tau)
&&=\sum_{m,n=0}^{\infty}J_{n}(m\alpha)[I^{\text{nd}}_{m}\sin(m\phi_0+m\Delta\mu\tau-n\omega_1\tau)+
(-1)^nI^{\text{d}}_{m}\cos(m\phi_0+m\Delta\mu\tau-n\omega_1\tau)]\nonumber\\
&&\ \ +\sum_{m=0}^{\infty}\sum_{n=1}^{\infty}J_{n}(m\alpha)[I^{\text{nd}}_{m}\sin(m\phi_0+m\Delta\mu\tau+n\omega_1\tau)
+I^{\text{d}}_{m}\cos(m\phi_0+m\Delta\mu\tau+n\omega_1\tau)].
\eeq
\end{widetext}
In the above, the second line in the right hand side always depends on time under $\Delta\mu,\omega_1>0$, and therefore is typically averaged out
in two-terminal setups of ultracold atomic gases.
However, the first line becomes time independent and gives a nonzero DC contribution if the following condition is satisfied:
\beq
\Delta\mu=\frac{n}{m}\omega_1.
\eeq
This is the condition that the Shapiro resonance occurs.
For instance, if 
\beq
\Delta\mu=\frac{\omega_1}{3},
\eeq
third-order harmonics of the AC current is dominantly contributing to the DC current, since the lower harmonics are averaged out.

\section{Concluding remarks} Our work elucidates the relation between the dissipative current and the superfluid character of the BEC. The QPC implements a coherent coupling of the superfluid component with the phonon modes, in the spirit of the celebrated Landau argument on superfluidity. Even though the zero temperature DC component does not carry any entropy, as demonstrated by the zero value of the Seebeck coefficient, it cannot be identified with a superfluid current. This also illustrates the difficulty in modeling the dynamics of superfluids in mesoscopic structures using effective two-fluid models: in particular QPCs behave very differently from super-leaks in helium, and the fountain effect is not expected \cite{Karpiuk:2012aa}.

Charge neutral fermionic superfluids also feature a gapless mode resulting from the $U(1)$ symmetry breaking. While we cannot extrapolate the bosonic results derived here to the fermionic case~\cite{uchino2020role}, Bogoliubov theory is adequate for the far molecular side of the BEC-BCS crossover. The $1/g$ dependence of the zero temperature contribution suggests that it should become less relevant as the system approaches the Feshbach resonance. This is confirmed by the good agreement between low temperature transport experiments at unitarity and MAR theory observed in \cite{husmann2015}. 

A direct generalization of our theory could describe spin transport in spinor BECs, or include spin-orbit coupling in the reservoirs and the contact. It also predicts coherent heat current oscillations at $2\Delta \mu$, encountered in superconductors \cite{Fornieri:2017aa}, but never studied so far with cold atoms, as well as the current noise spectrum. While current noise in the two-terminal setup has never been measured in quantum gas experiments, we expect it to become accessible in future generations of experimental setups~\cite{PhysRevA.98.063619}. 

\section*{Acknowledgement}
We thank Thierry Giamarchi and Martin Lebrat for careful reading of the manuscript and discussions, and also thank Masahito Ueda for discussions. S.U. is supported by JSPS KAKENHI Grant Number JP17K14366, Matsuo Foundation, and  Waseda University Grant for Special Research Projects (No. 2019C-461). J.P.B. acknowledges funding from the European Research Council (ERC) under the European Union's Horizon 2020 research and innovation program (grant agreement No 714309), the Sandoz Family Foundation-Monique de Meuron program for Academic Promotion, the SNSF (Project No. 184654) and EPFL.

\appendix

\section{Current expression}
Here we provide some details on  derivation of the current formula  (Eqs.~\eqref{eq:current}-\eqref{eq:even-current})
underlying the results of this work. The notations are those of the main text. Using Heisenberg equations of motion, particle and heat currents in the tunneling Hamiltonian can be expressed as
\beq
I_1(\tau)&&=-\frac{it}{\Omega}\sum_{\mathbf{p,k}}\langle b^{\dagger}_{\mathbf{p},R}(\tau)b_{\mathbf{k},L}(\tau)-
b^{\dagger}_{\mathbf{k},L}(\tau)b_{\mathbf{p},R}(\tau)\rangle,\\
I_2(\tau)&&=\frac{t}{\Omega}\sum_{\mathbf{p,k}}\Big\langle b^{\dagger}_{\mathbf{p},R}(\tau)\frac{d}{d\tau}b_{\mathbf{k},L}(\tau)+\frac{d}{d\tau}
b^{\dagger}_{\mathbf{k},L}(\tau)b_{\mathbf{p},R}(\tau)\Big\rangle,\nonumber\\
\eeq
where the field operator in the momentum space satisfying $\psi_{L(R)}(\mathbf{x})=\frac{1}{\sqrt{\Omega}} \sum_{\mathbf{k}} e^{i\mathbf{k}\cdot\mathbf{x}}b_{\mathbf{k},L(R)}$, and $\langle\cdots\rangle$ is the average under the total Hamiltonian with the tunneling term.
To go further, we rewrite the above expression as follows:
\beq
&&I_1(\tau)=2\text{Re}\Big[\hat{G}^<_{LR}(\tau,\tau)\hat{t}(\tau)\Big]_{11},\\
&&I_2(\tau)=
2\lim_{\tau'\to\tau}\frac{d}{d\tau}\text{Re}\Big[i\hat{G}^<_{LR}(\tau,\tau')\hat{t}(\tau') \Big]_{11},
\eeq
where 
\beq
&&\hat{G}^<_{LR}(\tau,\tau')\nonumber\\
&& =-\frac{i}{\Omega}\sum_{\mathbf{p,k}}
\begin{pmatrix}
 \langle b^{\dagger}_{\mathbf{p},R}(\tau') b_{\mathbf{k},L}(\tau)\rangle & \langle b_{\mathbf{p},R}(\tau') b_{\mathbf{k},L}(\tau)\rangle\\
\langle b^{\dagger}_{\mathbf{p},R}(\tau') b^{\dagger}_{\mathbf{k},L}(\tau)\rangle & \langle b_{\mathbf{p},R}(\tau') b^{\dagger}_{\mathbf{k},L}(\tau)\rangle
\end{pmatrix},
\eeq
\beq
&&\hat{t}(\tau)=t
\begin{pmatrix}
e^{-i\phi(\tau)} & 0\\
0 & e^{i\phi(\tau)}
\end{pmatrix},
\eeq
and we perform the gauge transformation in the tunneling Hamiltonian~\cite{mahan2013,PhysRevB.54.7366}, where
the chemical potential difference between the reservoirs is reflected in the phase factor of the tunneling amplitude.
Since it is difficult to evaluate the nonequilibrium average $\langle\cdots\rangle$ directly, 
we perform the expansion in $t$ by means of the Keldysh formalism, which allows us to
calculate the currents  in terms of Green's functions at equilibrium.
Indeed, by using the so-called Langreth rules~\cite{rammer2007}, we can obtain the following Dyson equation:
\beq
\hat{G}^<=(\hat{1}+\hat{G}^R\circ\hat{V})\circ\hat{g}^<\circ(\hat{1}+\hat{V}\circ\hat{G}^A),
\eeq
where $\hat{1}=\text{diag}(1,1)\delta(\tau-\tau')$,
$\hat{V}$ represents $\hat{t}$ or $\hat{t}^{\dagger}$, and $\hat{G}^{R(A)}$ is coupled retarded (advanced) Green's function obeying
\beq
\hat{G}^{R(A)}=\hat{g}^{R(A)}+\hat{g}^{R(A)}\circ\hat{V}\circ\hat{G}^{R(A)}.
\eeq
Here, we introduce uncoupled  retarded (advanced) Green's function $\hat{g}^{R(A)}$, which is evaluated at equilibrium. By using the low-frequency expansions in the Bogoliubov theory,
this is expressed in frequency space as
\begin{widetext}
\beq
\hat{g}^R_{\alpha\beta}(\omega)&&=-\frac{i}{\Omega}\sum_{\mathbf{k}}\int_{-\infty}^{\infty}d\tau e^{i\omega\tau}\theta(\tau)
\begin{pmatrix}
\langle[b_{\mathbf{k},\alpha}(\tau),b^{\dagger}_{\mathbf{k},\beta}(0)] \rangle_0 & \langle[b_{\mathbf{k},\alpha}(\tau),b_{\mathbf{-k},\beta}(0)] \rangle_0\\
\langle[b^{\dagger}_{\mathbf{-k},\alpha}(\tau),b^{\dagger}_{\mathbf{k},\beta}(0)] \rangle_0 & \langle[b^{\dagger}_{\mathbf{k},\alpha}(\tau),b_{\mathbf{k},\beta}(0)] \rangle_0
\end{pmatrix}
\nonumber\\
&&\approx\delta_{\alpha,\beta}
\begin{pmatrix}
-\frac{\mu_{\alpha}\omega_c}{2\pi^2c^3}-\frac{\omega_c\omega}{2\pi^2c_{\alpha}^3}+\frac{\mu_{\alpha}\omega^2}{2\pi^2\omega_cc_{\alpha}^3}
-i\Big[\frac{\mu_{\alpha}\omega}{4\pi c_{\alpha}^3}+\frac{\omega^2}{4\pi c_{\alpha}^3}\Big] &\frac{\mu_{\alpha}\omega_c}{2\pi^2c_{\alpha}^3}-\frac{\mu_{\alpha}\omega^2}{2\pi^2\omega_cc_{\alpha}^3}+i\frac{\mu_{\alpha}\omega}{4\pi c_{\alpha}^3}
\\ \frac{\mu_{\alpha}\omega_c}{2\pi^2c_{\alpha}^3}-\frac{\mu_{\alpha}\omega^2}{2\pi^2\omega_cc_{\alpha}^3}+i\frac{\mu_{\alpha}\omega}{4\pi c_{\alpha}^3} & -\frac{\mu_{\alpha}\omega_c}{2\pi^2c_{\alpha}^3}+\frac{\omega_c\omega}{2\pi^2c_{\alpha}^3}+\frac{\mu_{\alpha}\omega^2}{2\pi^2\omega_cc_{\alpha}^3}+i\Big[-\frac{\mu_{\alpha}\omega}{4\pi c_{\alpha}^3}+\frac{\omega^2}{4\pi c_{\alpha}^3}\Big]
\end{pmatrix},
\eeq
where $\alpha,\beta=$L or R, $\hat{g}^A(\omega)=[\hat{g}^{R}(\omega)]^{\dagger}$, and 
 $\omega_c$ is the cut-off frequency of the tunneling Hamiltonian and is chosen as the average chemical potential~\cite{PhysRevA.64.033610}.
Similarly, uncoupled lesser Green's function in frequency space is obtained as
\beq
\hat{g}^<_{\alpha\beta}(\omega)&&=-\frac{i}{\Omega}\sum_{\mathbf{k}}\int_{-\infty}^{\infty}d\tau e^{i\omega\tau}
\begin{pmatrix}
\langle b^{\dagger}_{\mathbf{k},\beta}(0) b_{\mathbf{k},\alpha}(\tau)\rangle_0 & \langle b_{\mathbf{-k},\beta}(0) b_{\mathbf{k},\alpha}(\tau)\rangle_0 \\
\langle b^{\dagger}_{\mathbf{k},\beta}(0) b^{\dagger}_{\mathbf{-k},\alpha}(\tau)\rangle_0  & \langle b_{\mathbf{k},\beta}(0) b^{\dagger}_{\mathbf{k},\alpha}(\tau)\rangle_0 
\end{pmatrix}
\nonumber\\
&&\approx-\frac{2\pi i\sqrt{\mu_{\alpha}\mu_{\beta}}}{g}\delta(\omega)
\begin{pmatrix}
1 &1 \\
1 & 1
\end{pmatrix}+i\delta_{\alpha,\beta}n_{\alpha}(\omega)
\begin{pmatrix}
-\Big[\frac{\mu_{\alpha}\omega}{2\pi c_{\alpha}^3}+\frac{\omega^2}{2\pi c_{\alpha}^3}\Big] & \frac{\mu_{\alpha}\omega }{2\pi c_{\alpha}^3}\\
\frac{\mu_{\alpha}\omega }{2\pi c_{\alpha}^3} & -\Big[\frac{\mu_{\alpha}\omega}{2\pi c_{\alpha}^3}-\frac{\omega^2}{2\pi c_{\alpha}^3}\Big]
\end{pmatrix},
\eeq
where
\beq
n_{\alpha}(\omega)=\frac{1}{e^{\omega/T_{\alpha}}-1}
\label{eq:bose-distribution}
\eeq
is the Bose distribution function.
The peculiar feature of lesser Green's function above  is  the presence of the contribution from BECs corresponding to the first term in the right-hand side of Eq.~(10).
This term is present even if $\alpha\ne\beta$, which leads to the presence of odd-order harmonics in the current. Indeed, by using uncoupled Green's functions,
we obtain
\beq
\hat{G}^{<,\text{even}}_{LR}&&=(\hat{1}+\hat{G}^R_{LR}\hat{t}) \circ \hat{g}^<_{LL}\circ\hat{t}^{\dagger}\hat{G}^A_{RR}
+\hat{G}^R_{LL}\hat{t}^{\dagger}\circ\hat{g}^<_{RR}\circ(\hat{1}+\hat{t}\hat{G}^A_{LR}),\\
\hat{G}^{<,\text{odd}}_{LR}&&=(\hat{1}+\hat{G}^R_{LR}\hat{t}) \circ \hat{g}^<_{LR}\circ(\hat{1}+\hat{t}\hat{G}^A_{LR})
+\hat{G}^R_{LL}\hat{t}^{\dagger}\circ\hat{g}^<_{RL}\circ\hat{t}^{\dagger}\hat{G}^A_{RR},
\eeq
where even (odd) in the superscript above represents the components of even (odd) harmonics in the current.
As is clear from the expressions above, $\hat{G}^{<,\text{odd}}_{LR}=\hat{0}$, since $\hat{g}^<_{LR}=\hat{g}^<_{RL}=\hat{0}$ in the fermionic system.
\end{widetext}
For efficient numerics, as in the case of superconductors, it is convenient to introduce the renormalized tunneling matrix~\eqref{eq:tunnel}.
By using this renormalized tunneling matrix,  the current formula shown in Sec.~II can be  obtained.

\section{Current expressions up to linear response}
We now derive several current expressions up to the linear response theory  
discussed in Sec. III.
In this case, the substitution $\hat{T}^{R(A)}\to\hat{t}$ is allowed, and analytic expressions are obtained. 

We first examine the current components linear in $t$.
The presence of such  components allows nonzero $I^{\text{odd}}_q(\tau)$.
By substituting  $\hat{T}^{R(A)}\to\hat{t}$ into Eq.~\eqref{eq:odd-current}, we obtain
\beq
&&I_{1,1}=0,\\
&&I_{1,-1}=-\frac{2it\sqrt{\mu_L\mu_R}}{g}.
\eeq
Thus, the particle current up to $t$ is given by $I_{1}(\tau)=-\frac{2t\sqrt{\mu_L\mu_R}}{g}\sin\phi(\tau)$.
 The similar calculation is allowed for the heat current. It is then straightforward to show $I_{2}(\tau)=0$, that is, absence of the Josephson heat
 current up to $t$.

We next examine the current components quadratic in $t$, which arise from $I^{\text{even}}_q(\tau)$ and include both DC and AC parts.
The DC part of the particle current is expressed as 
\beq
I_{1,0}&&=\int_{-\infty}^{\infty}\frac{t^2d\omega}{\pi}\text{Re}[\hat{g}^{R}_{LL,11}(\omega) \hat{g}^<_{RR,11}(\omega+\Delta\mu) \nonumber\\
&&\ \ +\hat{g}^<_{LL,11}(\omega)\hat{g}^A_{RR,11}(\omega+\Delta\mu) ]\nonumber\\
&&=I^{c}_{1,0}+I^{b}_{1,0},
\eeq
where we introduce 
\beq
I^{c}_{1,0}&&=\frac{t^2}{g}\Big[\mu_R\rho_L(-\Delta\mu)-\mu_L\rho_R(\Delta\mu) \Big],   \\
I^{b}_{1,0}&&=t^2\int_{-\infty}^{\infty}\frac{d\omega}{2\pi}\rho_{L}(\omega)\rho_R(\omega+\Delta\mu)\Big[n_L(\omega)-n_R(\omega+\Delta\mu) \Big],\nonumber\\
\eeq
with the density of state
\beq
\rho_{\alpha}(\omega)=-2\text{Im}[\hat{g}^R_{\alpha\alpha,11}(\omega)].
\eeq
Physically, $I^c_{1,0}$ describes  the contribution from the conversion process between condensate and phonon,
and $I^b_{1,0}$ describes the contribution from the tunneling process of phonons, which is
comparable to the tunneling current formula in fermionic quasiparticles~\cite{mahan2013}.
Based on the above expressions, we next look at the regime up to linear in $\Delta\mu$ and $\Delta T$, and determine
 transport coefficients in Onsager's matrix. 
 By using Eq.~\eqref{eq:bose-distribution}, $I^{c}_{1,0}$ is easily obtained as
 \beq
 I^c_{1,0}\approx \frac{t^2\mu^2}{\pi c^3g}\Delta\mu,
 \eeq
 where we introduce the average chemical potential $\mu$ and average sound velocity $c$.
 On the other hand, by using
 \beq
 &&n_L(\omega)-n_R(\omega+\Delta\mu)\approx \frac{\partial n(\omega)}{\partial\omega}\Delta\mu+\frac{\partial n(\omega)}{\partial T}\Delta T
 \nonumber\\
&&= \frac{1}{4T}\frac{\Delta\mu}{\sinh^2(\omega/(2T))}+\frac{\omega}{4T^2}\frac{\Delta T}{\sinh^2(\omega/(2T))},
 \eeq
 with the average temperature $T$,
$I^b_{1,0}$ is obtained as
 \beq
 I^{b}_{1,0}&&\approx L^b_{11}\Delta\mu+L_{12}\Delta T,
 \eeq
where 
\beq
L^b_{11}&&=\frac{t^2}{4T}\int_{-\infty}^{\infty}\frac{d\omega}{2\pi}\frac{\rho_{L}(\omega)\rho_R(\omega)}{\sinh^2(\omega/(2T))},\nonumber\\
             &&\approx\frac{t^2\mu^2T^2}{4\pi^3c^6}\int_{-\infty}^{\infty}dx\frac{x^2}{\sinh^2x},\nonumber\\
             &&=\frac{t^2\mu^2T^2}{12\pi c^6},\\
L_{12}&&=\frac{t^2}{4T^2}\int_{-\infty}^{\infty}\frac{d\omega}{2\pi}\frac{\omega\rho_{L}(\omega)\rho_R(\omega)}{\sinh^2(\omega/(2T))}\nonumber\\
          &&\approx\frac{2t^2\mu T^3 }{\pi^3c^6}\int_{-\infty}^{\infty}dx\frac{x^4}{\sinh^2x},\nonumber\\
          &&=\frac{2\pi t^2\mu T^3 }{15c^6}.
\eeq
To obtain above, we only keep the contributions with the leading temperature dependence. 
In $L^b_{11}$, such a contribution arises from a term 
$\propto \int_{-\infty}^{\infty}\frac{d\omega}{2\pi}\frac{\omega^2}{\sinh^2(\omega/(2T))}$.
Similarly, the leading order contribution in 
$L_{12}$ arises from a term $\propto \int_{-\infty}^{\infty}\frac{d\omega}{2\pi}\frac{\omega^4}{\sinh^2(\omega/(2T))}$~\footnote{We note
$\int_{-\infty}^{\infty}\frac{d\omega}{2\pi}\frac{\omega^3}{\sinh^2(\omega/(2T))}=0$}.

In addition, $L_{22}$ can be obtained from the DC part of the heat current at $\Delta\mu=0$, which is given by
\beq
I_{2,0}&&=\int_{-\infty}^{\infty}\frac{t^2\omega d\omega}{\pi}\text{Re}[\hat{g}^{R}_{LL,11}(\omega) \hat{g}^<_{RR,11}(\omega) 
+\hat{g}^<_{LL,11}(\omega)\hat{g}^A_{RR,11}(\omega) ]\nonumber\\
         &&=t^2\int_{-\infty}^{\infty}\frac{d\omega}{2\pi}\omega\rho_{L}(\omega)\rho_R(\omega)\Big[n_L(\omega)-n_R(\omega) \Big]\nonumber\\
         &&\approx L_{22}\Delta T.
\eeq
As in the case of $L^b_{11}$ and $L_{12}$, $L_{22}$ can be obtained as
\beq
L_{22}&&\approx\frac{t^2\mu^2T^3}{\pi^3 c^6}\int_{-\infty}^{\infty}dx\frac{x^4}{\sinh^2x}\nonumber\\
          &&=\frac{\pi t^2\mu^2T^3}{15 c^6}
\eeq

Calculation of the AC components is done in a similar manner.
At zero temperature, the AC components in the particle current are obtained as
\beq
I_{1,2}&&=0,\\
I_{1,-2}&&=-\frac{2i\mu_Rt^2}{g}\hat{g}^R_{LL,12}(\Delta\mu)-\frac{2i\mu_Lt^2}{g}\hat{g}^A_{RR,12}(-\Delta\mu).\nonumber\\
\eeq
Therefore, in the low-bias regime,
the harmonics proportional to $e^{\pm i2\phi(\tau)}$ is approximated as
\beq
\frac{t^2\mu^2\Delta\mu}{\pi g c^3}\cos(2\phi(\tau))-\frac{2t^2\mu^3}{\pi gc^3}\sin(2\phi(\tau)).
\eeq
The result above shows that the AC components quadratic in $t$ yield both dissipative and non-dissipative terms.
In addition, at the low bias, the amplitude of the dissipative term depends on $\Delta\mu$ and that of the non-dissipative term is independent of $\Delta\mu$.

\bibliographystyle{apsrev4-1}
%\bibliography{reference}

\begin{thebibliography}{64}%
\makeatletter
\providecommand \@ifxundefined [1]{%
 \@ifx{#1\undefined}
}%
\providecommand \@ifnum [1]{%
 \ifnum #1\expandafter \@firstoftwo
 \else \expandafter \@secondoftwo
 \fi
}%
\providecommand \@ifx [1]{%
 \ifx #1\expandafter \@firstoftwo
 \else \expandafter \@secondoftwo
 \fi
}%
\providecommand \natexlab [1]{#1}%
\providecommand \enquote  [1]{``#1''}%
\providecommand \bibnamefont  [1]{#1}%
\providecommand \bibfnamefont [1]{#1}%
\providecommand \citenamefont [1]{#1}%
\providecommand \href@noop [0]{\@secondoftwo}%
\providecommand \href [0]{\begingroup \@sanitize@url \@href}%
\providecommand \@href[1]{\@@startlink{#1}\@@href}%
\providecommand \@@href[1]{\endgroup#1\@@endlink}%
\providecommand \@sanitize@url [0]{\catcode `\\12\catcode `\$12\catcode
  `\&12\catcode `\#12\catcode `\^12\catcode `\_12\catcode `\%12\relax}%
\providecommand \@@startlink[1]{}%
\providecommand \@@endlink[0]{}%
\providecommand \url  [0]{\begingroup\@sanitize@url \@url }%
\providecommand \@url [1]{\endgroup\@href {#1}{\urlprefix }}%
\providecommand \urlprefix  [0]{URL }%
\providecommand \Eprint [0]{\href }%
\providecommand \doibase [0]{http://dx.doi.org/}%
\providecommand \selectlanguage [0]{\@gobble}%
\providecommand \bibinfo  [0]{\@secondoftwo}%
\providecommand \bibfield  [0]{\@secondoftwo}%
\providecommand \translation [1]{[#1]}%
\providecommand \BibitemOpen [0]{}%
\providecommand \bibitemStop [0]{}%
\providecommand \bibitemNoStop [0]{.\EOS\space}%
\providecommand \EOS [0]{\spacefactor3000\relax}%
\providecommand \BibitemShut  [1]{\csname bibitem#1\endcsname}%
\let\auto@bib@innerbib\@empty
%</preamble>
\bibitem [{\citenamefont {Nazarov}\ and\ \citenamefont
  {Blanter}(2009)}]{nazarov2009}%
  \BibitemOpen
  \bibfield  {author} {\bibinfo {author} {\bibfnamefont {Y.~V.}\ \bibnamefont
  {Nazarov}}\ and\ \bibinfo {author} {\bibfnamefont {Y.~M.}\ \bibnamefont
  {Blanter}},\ }\href@noop {} {\emph {\bibinfo {title} {Quantum transport:
  introduction to nanoscience}}}\ (\bibinfo  {publisher} {Cambridge University
  Press},\ \bibinfo {year} {2009})\BibitemShut {NoStop}%
\bibitem [{\citenamefont {Datta}(1997)}]{datta1997}%
  \BibitemOpen
  \bibfield  {author} {\bibinfo {author} {\bibfnamefont {S.}~\bibnamefont
  {Datta}},\ }\href@noop {} {\emph {\bibinfo {title} {Electronic transport in
  mesoscopic systems}}}\ (\bibinfo  {publisher} {Cambridge university press},\
  \bibinfo {year} {1997})\BibitemShut {NoStop}%
\bibitem [{\citenamefont {Blanter}\ and\ \citenamefont
  {B{\"u}ttiker}(2000)}]{blanter2000}%
  \BibitemOpen
  \bibfield  {author} {\bibinfo {author} {\bibfnamefont {Y.~M.}\ \bibnamefont
  {Blanter}}\ and\ \bibinfo {author} {\bibfnamefont {M.}~\bibnamefont
  {B{\"u}ttiker}},\ }\href@noop {} {\bibfield  {journal} {\bibinfo  {journal}
  {Physics reports}\ }\textbf {\bibinfo {volume} {336}},\ \bibinfo {pages} {1}
  (\bibinfo {year} {2000})}\BibitemShut {NoStop}%
\bibitem [{\citenamefont {Esposito}\ \emph {et~al.}(2009)\citenamefont
  {Esposito}, \citenamefont {Harbola},\ and\ \citenamefont
  {Mukamel}}]{RevModPhys.81.1665}%
  \BibitemOpen
  \bibfield  {author} {\bibinfo {author} {\bibfnamefont {M.}~\bibnamefont
  {Esposito}}, \bibinfo {author} {\bibfnamefont {U.}~\bibnamefont {Harbola}}, \
  and\ \bibinfo {author} {\bibfnamefont {S.}~\bibnamefont {Mukamel}},\ }\href
  {\doibase 10.1103/RevModPhys.81.1665} {\bibfield  {journal} {\bibinfo
  {journal} {Rev. Mod. Phys.}\ }\textbf {\bibinfo {volume} {81}},\ \bibinfo
  {pages} {1665} (\bibinfo {year} {2009})}\BibitemShut {NoStop}%
\bibitem [{\citenamefont {Chien}\ \emph {et~al.}(2015)\citenamefont {Chien},
  \citenamefont {Peotta},\ and\ \citenamefont {Di~Ventra}}]{Chien:2015ab}%
  \BibitemOpen
  \bibfield  {author} {\bibinfo {author} {\bibfnamefont {C.-C.}\ \bibnamefont
  {Chien}}, \bibinfo {author} {\bibfnamefont {S.}~\bibnamefont {Peotta}}, \
  and\ \bibinfo {author} {\bibfnamefont {M.}~\bibnamefont {Di~Ventra}},\ }\href
  {http://dx.doi.org/10.1038/nphys3531} {\bibfield  {journal} {\bibinfo
  {journal} {Nat Phys}\ }\textbf {\bibinfo {volume} {11}},\ \bibinfo {pages}
  {998} (\bibinfo {year} {2015})}\BibitemShut {NoStop}%
\bibitem [{\citenamefont {Krinner}\ \emph {et~al.}(2017)\citenamefont
  {Krinner}, \citenamefont {Esslinger},\ and\ \citenamefont
  {Brantut}}]{krinner2017}%
  \BibitemOpen
  \bibfield  {author} {\bibinfo {author} {\bibfnamefont {S.}~\bibnamefont
  {Krinner}}, \bibinfo {author} {\bibfnamefont {T.}~\bibnamefont {Esslinger}},
  \ and\ \bibinfo {author} {\bibfnamefont {J.-P.}\ \bibnamefont {Brantut}},\
  }\href@noop {} {\bibfield  {journal} {\bibinfo  {journal} {Journal of
  Physics: Condensed Matter}\ }\textbf {\bibinfo {volume} {29}},\ \bibinfo
  {pages} {343003} (\bibinfo {year} {2017})}\BibitemShut {NoStop}%
\bibitem [{\citenamefont {Krinner}\ \emph {et~al.}(2015)\citenamefont
  {Krinner}, \citenamefont {Stadler}, \citenamefont {Husmann}, \citenamefont
  {Brantut},\ and\ \citenamefont {Esslinger}}]{krinner2015}%
  \BibitemOpen
  \bibfield  {author} {\bibinfo {author} {\bibfnamefont {S.}~\bibnamefont
  {Krinner}}, \bibinfo {author} {\bibfnamefont {D.}~\bibnamefont {Stadler}},
  \bibinfo {author} {\bibfnamefont {D.}~\bibnamefont {Husmann}}, \bibinfo
  {author} {\bibfnamefont {J.-P.}\ \bibnamefont {Brantut}}, \ and\ \bibinfo
  {author} {\bibfnamefont {T.}~\bibnamefont {Esslinger}},\ }\href@noop {}
  {\bibfield  {journal} {\bibinfo  {journal} {Nature}\ }\textbf {\bibinfo
  {volume} {517}},\ \bibinfo {pages} {64} (\bibinfo {year} {2015})}\BibitemShut
  {NoStop}%
\bibitem [{\citenamefont {Seaman}\ \emph {et~al.}(2007)\citenamefont {Seaman},
  \citenamefont {Kr\"amer}, \citenamefont {Anderson},\ and\ \citenamefont
  {Holland}}]{Seaman:2007aa}%
  \BibitemOpen
  \bibfield  {author} {\bibinfo {author} {\bibfnamefont {B.~T.}\ \bibnamefont
  {Seaman}}, \bibinfo {author} {\bibfnamefont {M.}~\bibnamefont {Kr\"amer}},
  \bibinfo {author} {\bibfnamefont {D.~Z.}\ \bibnamefont {Anderson}}, \ and\
  \bibinfo {author} {\bibfnamefont {M.~J.}\ \bibnamefont {Holland}},\ }\href
  {\doibase 10.1103/PhysRevA.75.023615} {\bibfield  {journal} {\bibinfo
  {journal} {Phys. Rev. A}\ }\textbf {\bibinfo {volume} {75}},\ \bibinfo
  {pages} {023615} (\bibinfo {year} {2007})}\BibitemShut {NoStop}%
\bibitem [{\citenamefont {Ramanathan}\ \emph {et~al.}(2011)\citenamefont
  {Ramanathan}, \citenamefont {Wright}, \citenamefont {Muniz}, \citenamefont
  {Zelan}, \citenamefont {Hill}, \citenamefont {Lobb}, \citenamefont
  {Helmerson}, \citenamefont {Phillips},\ and\ \citenamefont
  {Campbell}}]{Ramanathan:2011aa}%
  \BibitemOpen
  \bibfield  {author} {\bibinfo {author} {\bibfnamefont {A.}~\bibnamefont
  {Ramanathan}}, \bibinfo {author} {\bibfnamefont {K.~C.}\ \bibnamefont
  {Wright}}, \bibinfo {author} {\bibfnamefont {S.~R.}\ \bibnamefont {Muniz}},
  \bibinfo {author} {\bibfnamefont {M.}~\bibnamefont {Zelan}}, \bibinfo
  {author} {\bibfnamefont {W.~T.}\ \bibnamefont {Hill}}, \bibinfo {author}
  {\bibfnamefont {C.~J.}\ \bibnamefont {Lobb}}, \bibinfo {author}
  {\bibfnamefont {K.}~\bibnamefont {Helmerson}}, \bibinfo {author}
  {\bibfnamefont {W.~D.}\ \bibnamefont {Phillips}}, \ and\ \bibinfo {author}
  {\bibfnamefont {G.~K.}\ \bibnamefont {Campbell}},\ }\href {\doibase
  10.1103/PhysRevLett.106.130401} {\bibfield  {journal} {\bibinfo  {journal}
  {Phys. Rev. Lett.}\ }\textbf {\bibinfo {volume} {106}},\ \bibinfo {pages}
  {130401} (\bibinfo {year} {2011})}\BibitemShut {NoStop}%
\bibitem [{\citenamefont {Jendrzejewski}\ \emph {et~al.}(2014)\citenamefont
  {Jendrzejewski}, \citenamefont {Eckel}, \citenamefont {Murray}, \citenamefont
  {Lanier}, \citenamefont {Edwards}, \citenamefont {Lobb},\ and\ \citenamefont
  {Campbell}}]{Jendrzejewski:2014aa}%
  \BibitemOpen
  \bibfield  {author} {\bibinfo {author} {\bibfnamefont {F.}~\bibnamefont
  {Jendrzejewski}}, \bibinfo {author} {\bibfnamefont {S.}~\bibnamefont
  {Eckel}}, \bibinfo {author} {\bibfnamefont {N.}~\bibnamefont {Murray}},
  \bibinfo {author} {\bibfnamefont {C.}~\bibnamefont {Lanier}}, \bibinfo
  {author} {\bibfnamefont {M.}~\bibnamefont {Edwards}}, \bibinfo {author}
  {\bibfnamefont {C.}~\bibnamefont {Lobb}}, \ and\ \bibinfo {author}
  {\bibfnamefont {G.}~\bibnamefont {Campbell}},\ }\href {\doibase
  10.1103/PhysRevLett.113.045305} {\bibfield  {journal} {\bibinfo  {journal}
  {Phys. Rev. Lett.}\ }\textbf {\bibinfo {volume} {113}},\ \bibinfo {pages}
  {045305} (\bibinfo {year} {2014})}\BibitemShut {NoStop}%
\bibitem [{\citenamefont {Ryu}\ and\ \citenamefont
  {Boshier}(2015)}]{Ryu:2015aa}%
  \BibitemOpen
  \bibfield  {author} {\bibinfo {author} {\bibfnamefont {C.}~\bibnamefont
  {Ryu}}\ and\ \bibinfo {author} {\bibfnamefont {M.~G.}\ \bibnamefont
  {Boshier}},\ }\bibfield  {booktitle} {\emph {\bibinfo {booktitle} {New
  Journal of Physics}},\ }\href {\doibase 10.1088/1367-2630/17/9/092002} {\
  \textbf {\bibinfo {volume} {17}},\ \bibinfo {pages} {092002} (\bibinfo {year}
  {2015})}\BibitemShut {NoStop}%
\bibitem [{\citenamefont {Bloch}\ \emph {et~al.}(2008)\citenamefont {Bloch},
  \citenamefont {Dalibard},\ and\ \citenamefont {Zwerger}}]{RevModPhys.80.885}%
  \BibitemOpen
  \bibfield  {author} {\bibinfo {author} {\bibfnamefont {I.}~\bibnamefont
  {Bloch}}, \bibinfo {author} {\bibfnamefont {J.}~\bibnamefont {Dalibard}}, \
  and\ \bibinfo {author} {\bibfnamefont {W.}~\bibnamefont {Zwerger}},\ }\href
  {\doibase 10.1103/RevModPhys.80.885} {\bibfield  {journal} {\bibinfo
  {journal} {Rev. Mod. Phys.}\ }\textbf {\bibinfo {volume} {80}},\ \bibinfo
  {pages} {885} (\bibinfo {year} {2008})}\BibitemShut {NoStop}%
\bibitem [{\citenamefont {Gutman}\ \emph {et~al.}(2012)\citenamefont {Gutman},
  \citenamefont {Gefen},\ and\ \citenamefont {Mirlin}}]{PhysRevB.85.125102}%
  \BibitemOpen
  \bibfield  {author} {\bibinfo {author} {\bibfnamefont {D.~B.}\ \bibnamefont
  {Gutman}}, \bibinfo {author} {\bibfnamefont {Y.}~\bibnamefont {Gefen}}, \
  and\ \bibinfo {author} {\bibfnamefont {A.~D.}\ \bibnamefont {Mirlin}},\
  }\href {\doibase 10.1103/PhysRevB.85.125102} {\bibfield  {journal} {\bibinfo
  {journal} {Phys. Rev. B}\ }\textbf {\bibinfo {volume} {85}},\ \bibinfo
  {pages} {125102} (\bibinfo {year} {2012})}\BibitemShut {NoStop}%
\bibitem [{\citenamefont {Chien}\ \emph {et~al.}(2014)\citenamefont {Chien},
  \citenamefont {Di~Ventra},\ and\ \citenamefont
  {Zwolak}}]{PhysRevA.90.023624}%
  \BibitemOpen
  \bibfield  {author} {\bibinfo {author} {\bibfnamefont {C.-C.}\ \bibnamefont
  {Chien}}, \bibinfo {author} {\bibfnamefont {M.}~\bibnamefont {Di~Ventra}}, \
  and\ \bibinfo {author} {\bibfnamefont {M.}~\bibnamefont {Zwolak}},\ }\href
  {\doibase 10.1103/PhysRevA.90.023624} {\bibfield  {journal} {\bibinfo
  {journal} {Phys. Rev. A}\ }\textbf {\bibinfo {volume} {90}},\ \bibinfo
  {pages} {023624} (\bibinfo {year} {2014})}\BibitemShut {NoStop}%
\bibitem [{\citenamefont {Kolovsky}\ \emph {et~al.}(2018)\citenamefont
  {Kolovsky}, \citenamefont {Denis},\ and\ \citenamefont
  {Wimberger}}]{PhysRevA.98.043623}%
  \BibitemOpen
  \bibfield  {author} {\bibinfo {author} {\bibfnamefont {A.~R.}\ \bibnamefont
  {Kolovsky}}, \bibinfo {author} {\bibfnamefont {Z.}~\bibnamefont {Denis}}, \
  and\ \bibinfo {author} {\bibfnamefont {S.}~\bibnamefont {Wimberger}},\ }\href
  {\doibase 10.1103/PhysRevA.98.043623} {\bibfield  {journal} {\bibinfo
  {journal} {Phys. Rev. A}\ }\textbf {\bibinfo {volume} {98}},\ \bibinfo
  {pages} {043623} (\bibinfo {year} {2018})}\BibitemShut {NoStop}%
\bibitem [{\citenamefont {Altland}\ and\ \citenamefont
  {Simons}(2010)}]{altland2010}%
  \BibitemOpen
  \bibfield  {author} {\bibinfo {author} {\bibfnamefont {A.}~\bibnamefont
  {Altland}}\ and\ \bibinfo {author} {\bibfnamefont {B.~D.}\ \bibnamefont
  {Simons}},\ }\href@noop {} {\emph {\bibinfo {title} {Condensed matter field
  theory}}}\ (\bibinfo  {publisher} {Cambridge University Press},\ \bibinfo
  {year} {2010})\BibitemShut {NoStop}%
\bibitem [{\citenamefont {Nagaosa}(2013)}]{nagaosa2013}%
  \BibitemOpen
  \bibfield  {author} {\bibinfo {author} {\bibfnamefont {N.}~\bibnamefont
  {Nagaosa}},\ }\href@noop {} {\emph {\bibinfo {title} {Quantum field theory in
  condensed matter physics}}}\ (\bibinfo  {publisher} {Springer Science \&
  Business Media},\ \bibinfo {year} {2013})\BibitemShut {NoStop}%
\bibitem [{\citenamefont {Zagoskin}(1998)}]{Zagoskin:1998aa}%
  \BibitemOpen
  \bibfield  {author} {\bibinfo {author} {\bibfnamefont {A.~M.}\ \bibnamefont
  {Zagoskin}},\ }\href@noop {} {\emph {\bibinfo {title} {Quantum theory of
  many-body systems}}},\ Vol.\ \bibinfo {volume} {174}\ (\bibinfo  {publisher}
  {Springer},\ \bibinfo {year} {1998})\BibitemShut {NoStop}%
\bibitem [{\citenamefont {Averin}\ and\ \citenamefont
  {Bardas}(1995)}]{PhysRevLett.75.1831}%
  \BibitemOpen
  \bibfield  {author} {\bibinfo {author} {\bibfnamefont {D.}~\bibnamefont
  {Averin}}\ and\ \bibinfo {author} {\bibfnamefont {A.}~\bibnamefont
  {Bardas}},\ }\href {\doibase 10.1103/PhysRevLett.75.1831} {\bibfield
  {journal} {\bibinfo  {journal} {Phys. Rev. Lett.}\ }\textbf {\bibinfo
  {volume} {75}},\ \bibinfo {pages} {1831} (\bibinfo {year}
  {1995})}\BibitemShut {NoStop}%
\bibitem [{\citenamefont {Eckel}\ \emph {et~al.}(2016)\citenamefont {Eckel},
  \citenamefont {Lee}, \citenamefont {Jendrzejewski}, \citenamefont {Lobb},
  \citenamefont {Campbell},\ and\ \citenamefont {Hill}}]{PhysRevA.93.063619}%
  \BibitemOpen
  \bibfield  {author} {\bibinfo {author} {\bibfnamefont {S.}~\bibnamefont
  {Eckel}}, \bibinfo {author} {\bibfnamefont {J.~G.}\ \bibnamefont {Lee}},
  \bibinfo {author} {\bibfnamefont {F.}~\bibnamefont {Jendrzejewski}}, \bibinfo
  {author} {\bibfnamefont {C.~J.}\ \bibnamefont {Lobb}}, \bibinfo {author}
  {\bibfnamefont {G.~K.}\ \bibnamefont {Campbell}}, \ and\ \bibinfo {author}
  {\bibfnamefont {W.~T.}\ \bibnamefont {Hill}},\ }\href {\doibase
  10.1103/PhysRevA.93.063619} {\bibfield  {journal} {\bibinfo  {journal} {Phys.
  Rev. A}\ }\textbf {\bibinfo {volume} {93}},\ \bibinfo {pages} {063619}
  (\bibinfo {year} {2016})}\BibitemShut {NoStop}%
\bibitem [{\citenamefont {Duc}\ \emph {et~al.}(2015)\citenamefont {Duc},
  \citenamefont {Savard}, \citenamefont {Petrescu}, \citenamefont {Rosenow},
  \citenamefont {Del~Maestro},\ and\ \citenamefont {Gervais}}]{Duc:2015aa}%
  \BibitemOpen
  \bibfield  {author} {\bibinfo {author} {\bibfnamefont {P.-F.}\ \bibnamefont
  {Duc}}, \bibinfo {author} {\bibfnamefont {M.}~\bibnamefont {Savard}},
  \bibinfo {author} {\bibfnamefont {M.}~\bibnamefont {Petrescu}}, \bibinfo
  {author} {\bibfnamefont {B.}~\bibnamefont {Rosenow}}, \bibinfo {author}
  {\bibfnamefont {A.}~\bibnamefont {Del~Maestro}}, \ and\ \bibinfo {author}
  {\bibfnamefont {G.}~\bibnamefont {Gervais}},\ }\href {\doibase
  10.1126/sciadv.1400222} {\bibfield  {journal} {\bibinfo  {journal} {Science
  Advances}\ }\textbf {\bibinfo {volume} {1}} (\bibinfo {year} {2015}),\
  10.1126/sciadv.1400222}\BibitemShut {NoStop}%
\bibitem [{\citenamefont {Albiez}\ \emph {et~al.}(2005)\citenamefont {Albiez},
  \citenamefont {Gati}, \citenamefont {F\"olling}, \citenamefont {Hunsmann},
  \citenamefont {Cristiani},\ and\ \citenamefont
  {Oberthaler}}]{PhysRevLett.95.010402}%
  \BibitemOpen
  \bibfield  {author} {\bibinfo {author} {\bibfnamefont {M.}~\bibnamefont
  {Albiez}}, \bibinfo {author} {\bibfnamefont {R.}~\bibnamefont {Gati}},
  \bibinfo {author} {\bibfnamefont {J.}~\bibnamefont {F\"olling}}, \bibinfo
  {author} {\bibfnamefont {S.}~\bibnamefont {Hunsmann}}, \bibinfo {author}
  {\bibfnamefont {M.}~\bibnamefont {Cristiani}}, \ and\ \bibinfo {author}
  {\bibfnamefont {M.~K.}\ \bibnamefont {Oberthaler}},\ }\href {\doibase
  10.1103/PhysRevLett.95.010402} {\bibfield  {journal} {\bibinfo  {journal}
  {Phys. Rev. Lett.}\ }\textbf {\bibinfo {volume} {95}},\ \bibinfo {pages}
  {010402} (\bibinfo {year} {2005})}\BibitemShut {NoStop}%
\bibitem [{\citenamefont {Levy}(2007)}]{levy2007}%
  \BibitemOpen
  \bibfield  {author} {\bibinfo {author} {\bibfnamefont {S.}~\bibnamefont
  {Levy}},\ }\href@noop {} {\bibfield  {journal} {\bibinfo  {journal} {Nature
  (London)}\ }\textbf {\bibinfo {volume} {449}},\ \bibinfo {pages} {579}
  (\bibinfo {year} {2007})}\BibitemShut {NoStop}%
\bibitem [{\citenamefont {LeBlanc}\ \emph {et~al.}(2011)\citenamefont
  {LeBlanc}, \citenamefont {Bardon}, \citenamefont {McKeever}, \citenamefont
  {Extavour}, \citenamefont {Jervis}, \citenamefont {Thywissen}, \citenamefont
  {Piazza},\ and\ \citenamefont {Smerzi}}]{PhysRevLett.106.025302}%
  \BibitemOpen
  \bibfield  {author} {\bibinfo {author} {\bibfnamefont {L.~J.}\ \bibnamefont
  {LeBlanc}}, \bibinfo {author} {\bibfnamefont {A.~B.}\ \bibnamefont {Bardon}},
  \bibinfo {author} {\bibfnamefont {J.}~\bibnamefont {McKeever}}, \bibinfo
  {author} {\bibfnamefont {M.~H.~T.}\ \bibnamefont {Extavour}}, \bibinfo
  {author} {\bibfnamefont {D.}~\bibnamefont {Jervis}}, \bibinfo {author}
  {\bibfnamefont {J.~H.}\ \bibnamefont {Thywissen}}, \bibinfo {author}
  {\bibfnamefont {F.}~\bibnamefont {Piazza}}, \ and\ \bibinfo {author}
  {\bibfnamefont {A.}~\bibnamefont {Smerzi}},\ }\href {\doibase
  10.1103/PhysRevLett.106.025302} {\bibfield  {journal} {\bibinfo  {journal}
  {Phys. Rev. Lett.}\ }\textbf {\bibinfo {volume} {106}},\ \bibinfo {pages}
  {025302} (\bibinfo {year} {2011})}\BibitemShut {NoStop}%
\bibitem [{\citenamefont {Xhani}\ \emph {et~al.}(2019)\citenamefont {Xhani},
  \citenamefont {Neri}, \citenamefont {Galantucci}, \citenamefont {Scazza},
  \citenamefont {Burchianti}, \citenamefont {Lee}, \citenamefont {Barenghi},
  \citenamefont {Trombettoni}, \citenamefont {Inguscio},\ and\ \citenamefont
  {Zaccanti}}]{Xhani:2019aa}%
  \BibitemOpen
  \bibfield  {author} {\bibinfo {author} {\bibfnamefont {K.}~\bibnamefont
  {Xhani}}, \bibinfo {author} {\bibfnamefont {E.}~\bibnamefont {Neri}},
  \bibinfo {author} {\bibfnamefont {L.}~\bibnamefont {Galantucci}}, \bibinfo
  {author} {\bibfnamefont {F.}~\bibnamefont {Scazza}}, \bibinfo {author}
  {\bibfnamefont {A.}~\bibnamefont {Burchianti}}, \bibinfo {author}
  {\bibfnamefont {K.-L.}\ \bibnamefont {Lee}}, \bibinfo {author} {\bibfnamefont
  {C.}~\bibnamefont {Barenghi}}, \bibinfo {author} {\bibfnamefont
  {A.}~\bibnamefont {Trombettoni}}, \bibinfo {author} {\bibfnamefont
  {M.}~\bibnamefont {Inguscio}}, \ and\ \bibinfo {author} {\bibfnamefont
  {M.}~\bibnamefont {Zaccanti}},\ }\href@noop {} {\bibfield  {journal}
  {\bibinfo  {journal} {arXiv preprint arXiv:1905.08893}\ } (\bibinfo {year}
  {2019})}\BibitemShut {NoStop}%
\bibitem [{\citenamefont {Cuevas}\ \emph {et~al.}(1996)\citenamefont {Cuevas},
  \citenamefont {Mart\'{\i}n-Rodero},\ and\ \citenamefont
  {Yeyati}}]{PhysRevB.54.7366}%
  \BibitemOpen
  \bibfield  {author} {\bibinfo {author} {\bibfnamefont {J.~C.}\ \bibnamefont
  {Cuevas}}, \bibinfo {author} {\bibfnamefont {A.}~\bibnamefont
  {Mart\'{\i}n-Rodero}}, \ and\ \bibinfo {author} {\bibfnamefont {A.~L.}\
  \bibnamefont {Yeyati}},\ }\href {\doibase 10.1103/PhysRevB.54.7366}
  {\bibfield  {journal} {\bibinfo  {journal} {Phys. Rev. B}\ }\textbf {\bibinfo
  {volume} {54}},\ \bibinfo {pages} {7366} (\bibinfo {year}
  {1996})}\BibitemShut {NoStop}%
\bibitem [{\citenamefont {Bolech}\ and\ \citenamefont
  {Giamarchi}(2004)}]{PhysRevLett.92.127001}%
  \BibitemOpen
  \bibfield  {author} {\bibinfo {author} {\bibfnamefont {C.~J.}\ \bibnamefont
  {Bolech}}\ and\ \bibinfo {author} {\bibfnamefont {T.}~\bibnamefont
  {Giamarchi}},\ }\href {\doibase 10.1103/PhysRevLett.92.127001} {\bibfield
  {journal} {\bibinfo  {journal} {Phys. Rev. Lett.}\ }\textbf {\bibinfo
  {volume} {92}},\ \bibinfo {pages} {127001} (\bibinfo {year}
  {2004})}\BibitemShut {NoStop}%
\bibitem [{\citenamefont {Bolech}\ and\ \citenamefont
  {Giamarchi}(2005)}]{PhysRevB.71.024517}%
  \BibitemOpen
  \bibfield  {author} {\bibinfo {author} {\bibfnamefont {C.~J.}\ \bibnamefont
  {Bolech}}\ and\ \bibinfo {author} {\bibfnamefont {T.}~\bibnamefont
  {Giamarchi}},\ }\href {\doibase 10.1103/PhysRevB.71.024517} {\bibfield
  {journal} {\bibinfo  {journal} {Phys. Rev. B}\ }\textbf {\bibinfo {volume}
  {71}},\ \bibinfo {pages} {024517} (\bibinfo {year} {2005})}\BibitemShut
  {NoStop}%
\bibitem [{\citenamefont {Uchino}\ and\ \citenamefont
  {Ueda}(2017)}]{PhysRevLett.118.105303}%
  \BibitemOpen
  \bibfield  {author} {\bibinfo {author} {\bibfnamefont {S.}~\bibnamefont
  {Uchino}}\ and\ \bibinfo {author} {\bibfnamefont {M.}~\bibnamefont {Ueda}},\
  }\href {\doibase 10.1103/PhysRevLett.118.105303} {\bibfield  {journal}
  {\bibinfo  {journal} {Phys. Rev. Lett.}\ }\textbf {\bibinfo {volume} {118}},\
  \bibinfo {pages} {105303} (\bibinfo {year} {2017})}\BibitemShut {NoStop}%
\bibitem [{\citenamefont {Yao}\ \emph {et~al.}(2018)\citenamefont {Yao},
  \citenamefont {Liu}, \citenamefont {Sun},\ and\ \citenamefont
  {Zhai}}]{PhysRevA.98.041601}%
  \BibitemOpen
  \bibfield  {author} {\bibinfo {author} {\bibfnamefont {J.}~\bibnamefont
  {Yao}}, \bibinfo {author} {\bibfnamefont {B.}~\bibnamefont {Liu}}, \bibinfo
  {author} {\bibfnamefont {M.}~\bibnamefont {Sun}}, \ and\ \bibinfo {author}
  {\bibfnamefont {H.}~\bibnamefont {Zhai}},\ }\href {\doibase
  10.1103/PhysRevA.98.041601} {\bibfield  {journal} {\bibinfo  {journal} {Phys.
  Rev. A}\ }\textbf {\bibinfo {volume} {98}},\ \bibinfo {pages} {041601}
  (\bibinfo {year} {2018})}\BibitemShut {NoStop}%
\bibitem [{\citenamefont {Han}\ \emph {et~al.}(2018)\citenamefont {Han},
  \citenamefont {Liu},\ and\ \citenamefont {Hu}}]{han2018}%
  \BibitemOpen
  \bibfield  {author} {\bibinfo {author} {\bibfnamefont {X.}~\bibnamefont
  {Han}}, \bibinfo {author} {\bibfnamefont {B.}~\bibnamefont {Liu}}, \ and\
  \bibinfo {author} {\bibfnamefont {J.}~\bibnamefont {Hu}},\ }\href@noop {}
  {\bibfield  {journal} {\bibinfo  {journal} {arXiv preprint arXiv:1806.09805}\
  } (\bibinfo {year} {2018})}\BibitemShut {NoStop}%
\bibitem [{\citenamefont {Andersen}(2004)}]{RevModPhys.76.599}%
  \BibitemOpen
  \bibfield  {author} {\bibinfo {author} {\bibfnamefont {J.~O.}\ \bibnamefont
  {Andersen}},\ }\href {\doibase 10.1103/RevModPhys.76.599} {\bibfield
  {journal} {\bibinfo  {journal} {Rev. Mod. Phys.}\ }\textbf {\bibinfo {volume}
  {76}},\ \bibinfo {pages} {599} (\bibinfo {year} {2004})}\BibitemShut
  {NoStop}%
\bibitem [{\citenamefont {Pitaevskii}\ and\ \citenamefont
  {Stringari}(2016)}]{pitaevskii2016}%
  \BibitemOpen
  \bibfield  {author} {\bibinfo {author} {\bibfnamefont {L.}~\bibnamefont
  {Pitaevskii}}\ and\ \bibinfo {author} {\bibfnamefont {S.}~\bibnamefont
  {Stringari}},\ }\href@noop {} {\emph {\bibinfo {title} {Bose-Einstein
  condensation and superfluidity}}},\ Vol.\ \bibinfo {volume} {164}\ (\bibinfo
  {publisher} {Oxford University Press},\ \bibinfo {year} {2016})\BibitemShut
  {NoStop}%
\bibitem [{\citenamefont {Rammer}(2007)}]{rammer2007}%
  \BibitemOpen
  \bibfield  {author} {\bibinfo {author} {\bibfnamefont {J.}~\bibnamefont
  {Rammer}},\ }\href@noop {} {\emph {\bibinfo {title} {Quantum field theory of
  non-equilibrium states}}}\ (\bibinfo  {publisher} {Cambridge University
  Press},\ \bibinfo {year} {2007})\BibitemShut {NoStop}%
\bibitem [{\citenamefont {Kamenev}(2011)}]{kamenev2011}%
  \BibitemOpen
  \bibfield  {author} {\bibinfo {author} {\bibfnamefont {A.}~\bibnamefont
  {Kamenev}},\ }\href@noop {} {\emph {\bibinfo {title} {Field theory of
  non-equilibrium systems}}}\ (\bibinfo  {publisher} {Cambridge University
  Press},\ \bibinfo {year} {2011})\BibitemShut {NoStop}%
\bibitem [{Note1()}]{Note1}%
  \BibitemOpen
  \bibinfo {note} {We note that transport coefficients depend on biases in
  non-Ohmic cases, where the Lorenz number cannot solely be expressed in terms
  of fundamental constants.}\BibitemShut {Stop}%
\bibitem [{\citenamefont {Gaunt}\ \emph {et~al.}(2013)\citenamefont {Gaunt},
  \citenamefont {Schmidutz}, \citenamefont {Gotlibovych}, \citenamefont
  {Smith},\ and\ \citenamefont {Hadzibabic}}]{PhysRevLett.110.200406}%
  \BibitemOpen
  \bibfield  {author} {\bibinfo {author} {\bibfnamefont {A.~L.}\ \bibnamefont
  {Gaunt}}, \bibinfo {author} {\bibfnamefont {T.~F.}\ \bibnamefont
  {Schmidutz}}, \bibinfo {author} {\bibfnamefont {I.}~\bibnamefont
  {Gotlibovych}}, \bibinfo {author} {\bibfnamefont {R.~P.}\ \bibnamefont
  {Smith}}, \ and\ \bibinfo {author} {\bibfnamefont {Z.}~\bibnamefont
  {Hadzibabic}},\ }\href {\doibase 10.1103/PhysRevLett.110.200406} {\bibfield
  {journal} {\bibinfo  {journal} {Phys. Rev. Lett.}\ }\textbf {\bibinfo
  {volume} {110}},\ \bibinfo {pages} {200406} (\bibinfo {year}
  {2013})}\BibitemShut {NoStop}%
\bibitem [{\citenamefont {Likharev}(1979)}]{RevModPhys.51.101}%
  \BibitemOpen
  \bibfield  {author} {\bibinfo {author} {\bibfnamefont {K.~K.}\ \bibnamefont
  {Likharev}},\ }\href {\doibase 10.1103/RevModPhys.51.101} {\bibfield
  {journal} {\bibinfo  {journal} {Rev. Mod. Phys.}\ }\textbf {\bibinfo {volume}
  {51}},\ \bibinfo {pages} {101} (\bibinfo {year} {1979})}\BibitemShut
  {NoStop}%
\bibitem [{\citenamefont {Mahan}(2013)}]{mahan2013}%
  \BibitemOpen
  \bibfield  {author} {\bibinfo {author} {\bibfnamefont {G.~D.}\ \bibnamefont
  {Mahan}},\ }\href@noop {} {\emph {\bibinfo {title} {Many-particle physics}}}\
  (\bibinfo  {publisher} {Springer Science \& Business Media},\ \bibinfo {year}
  {2013})\BibitemShut {NoStop}%
\bibitem [{\citenamefont {Berthod}\ and\ \citenamefont
  {Giamarchi}(2011)}]{PhysRevB.84.155414}%
  \BibitemOpen
  \bibfield  {author} {\bibinfo {author} {\bibfnamefont {C.}~\bibnamefont
  {Berthod}}\ and\ \bibinfo {author} {\bibfnamefont {T.}~\bibnamefont
  {Giamarchi}},\ }\href {\doibase 10.1103/PhysRevB.84.155414} {\bibfield
  {journal} {\bibinfo  {journal} {Phys. Rev. B}\ }\textbf {\bibinfo {volume}
  {84}},\ \bibinfo {pages} {155414} (\bibinfo {year} {2011})}\BibitemShut
  {NoStop}%
\bibitem [{\citenamefont {Fetter}\ and\ \citenamefont
  {Walecka}(2012)}]{fetter2012}%
  \BibitemOpen
  \bibfield  {author} {\bibinfo {author} {\bibfnamefont {A.~L.}\ \bibnamefont
  {Fetter}}\ and\ \bibinfo {author} {\bibfnamefont {J.~D.}\ \bibnamefont
  {Walecka}},\ }\href@noop {} {\emph {\bibinfo {title} {Quantum theory of
  many-particle systems}}}\ (\bibinfo  {publisher} {Courier Corporation},\
  \bibinfo {year} {2012})\BibitemShut {NoStop}%
\bibitem [{Note2()}]{Note2}%
  \BibitemOpen
  \bibinfo {note} {The c-number part of the field operator does not evolve in
  time and so is that of lesser Green's function~\cite
  {fetter2012}.}\BibitemShut {Stop}%
\bibitem [{\citenamefont {Meier}\ and\ \citenamefont
  {Zwerger}(2001)}]{PhysRevA.64.033610}%
  \BibitemOpen
  \bibfield  {author} {\bibinfo {author} {\bibfnamefont {F.}~\bibnamefont
  {Meier}}\ and\ \bibinfo {author} {\bibfnamefont {W.}~\bibnamefont
  {Zwerger}},\ }\href {\doibase 10.1103/PhysRevA.64.033610} {\bibfield
  {journal} {\bibinfo  {journal} {Phys. Rev. A}\ }\textbf {\bibinfo {volume}
  {64}},\ \bibinfo {pages} {033610} (\bibinfo {year} {2001})}\BibitemShut
  {NoStop}%
\bibitem [{\citenamefont {Husmann}\ \emph {et~al.}(2015)\citenamefont
  {Husmann}, \citenamefont {Uchino}, \citenamefont {Krinner}, \citenamefont
  {Lebrat}, \citenamefont {Giamarchi}, \citenamefont {Esslinger},\ and\
  \citenamefont {Brantut}}]{husmann2015}%
  \BibitemOpen
  \bibfield  {author} {\bibinfo {author} {\bibfnamefont {D.}~\bibnamefont
  {Husmann}}, \bibinfo {author} {\bibfnamefont {S.}~\bibnamefont {Uchino}},
  \bibinfo {author} {\bibfnamefont {S.}~\bibnamefont {Krinner}}, \bibinfo
  {author} {\bibfnamefont {M.}~\bibnamefont {Lebrat}}, \bibinfo {author}
  {\bibfnamefont {T.}~\bibnamefont {Giamarchi}}, \bibinfo {author}
  {\bibfnamefont {T.}~\bibnamefont {Esslinger}}, \ and\ \bibinfo {author}
  {\bibfnamefont {J.-P.}\ \bibnamefont {Brantut}},\ }\href@noop {} {\bibfield
  {journal} {\bibinfo  {journal} {Science}\ }\textbf {\bibinfo {volume}
  {350}},\ \bibinfo {pages} {1498} (\bibinfo {year} {2015})}\BibitemShut
  {NoStop}%
\bibitem [{\citenamefont {Giamarchi}(2004)}]{Giamarchi:2004aa}%
  \BibitemOpen
  \bibfield  {author} {\bibinfo {author} {\bibfnamefont {T.}~\bibnamefont
  {Giamarchi}},\ }\href@noop {} {\emph {\bibinfo {title} {Quantum Physics in
  One Dimension}}}\ (\bibinfo  {publisher} {Oxford University Press},\ \bibinfo
  {year} {2004})\BibitemShut {NoStop}%
\bibitem [{\citenamefont {van Houten}\ and\ \citenamefont
  {Beenakker}(1996)}]{Beenakker}%
  \BibitemOpen
  \bibfield  {author} {\bibinfo {author} {\bibfnamefont {H.}~\bibnamefont {van
  Houten}}\ and\ \bibinfo {author} {\bibfnamefont {C.~W.~J.}\ \bibnamefont
  {Beenakker}},\ }\href@noop {} {\bibfield  {journal} {\bibinfo  {journal}
  {Physics Today}\ }\textbf {\bibinfo {volume} {49}},\ \bibinfo {pages} {22}
  (\bibinfo {year} {1996})}\BibitemShut {NoStop}%
\bibitem [{\citenamefont {Papoular}\ \emph {et~al.}(2014)\citenamefont
  {Papoular}, \citenamefont {Pitaevskii},\ and\ \citenamefont
  {Stringari}}]{PhysRevLett.113.170601}%
  \BibitemOpen
  \bibfield  {author} {\bibinfo {author} {\bibfnamefont {D.~J.}\ \bibnamefont
  {Papoular}}, \bibinfo {author} {\bibfnamefont {L.~P.}\ \bibnamefont
  {Pitaevskii}}, \ and\ \bibinfo {author} {\bibfnamefont {S.}~\bibnamefont
  {Stringari}},\ }\href {\doibase 10.1103/PhysRevLett.113.170601} {\bibfield
  {journal} {\bibinfo  {journal} {Phys. Rev. Lett.}\ }\textbf {\bibinfo
  {volume} {113}},\ \bibinfo {pages} {170601} (\bibinfo {year}
  {2014})}\BibitemShut {NoStop}%
\bibitem [{\citenamefont {Ashcroft}\ and\ \citenamefont
  {Mermin}(1976)}]{ashcroft}%
  \BibitemOpen
  \bibfield  {author} {\bibinfo {author} {\bibfnamefont {N.~W.}\ \bibnamefont
  {Ashcroft}}\ and\ \bibinfo {author} {\bibfnamefont {N.~D.}\ \bibnamefont
  {Mermin}},\ }\href@noop {} {\emph {\bibinfo {title} {Solid State Physics}}}\
  (\bibinfo  {publisher} {Thomson Lerning},\ \bibinfo {year}
  {1976})\BibitemShut {NoStop}%
\bibitem [{\citenamefont {Nakata}\ \emph {et~al.}(2015)\citenamefont {Nakata},
  \citenamefont {Simon},\ and\ \citenamefont {Loss}}]{PhysRevB.92.134425}%
  \BibitemOpen
  \bibfield  {author} {\bibinfo {author} {\bibfnamefont {K.}~\bibnamefont
  {Nakata}}, \bibinfo {author} {\bibfnamefont {P.}~\bibnamefont {Simon}}, \
  and\ \bibinfo {author} {\bibfnamefont {D.}~\bibnamefont {Loss}},\ }\href
  {\doibase 10.1103/PhysRevB.92.134425} {\bibfield  {journal} {\bibinfo
  {journal} {Phys. Rev. B}\ }\textbf {\bibinfo {volume} {92}},\ \bibinfo
  {pages} {134425} (\bibinfo {year} {2015})}\BibitemShut {NoStop}%
\bibitem [{\citenamefont {Nakata}\ \emph {et~al.}(2018)\citenamefont {Nakata},
  \citenamefont {Ohnuma},\ and\ \citenamefont {Matsuo}}]{PhysRevB.98.094430}%
  \BibitemOpen
  \bibfield  {author} {\bibinfo {author} {\bibfnamefont {K.}~\bibnamefont
  {Nakata}}, \bibinfo {author} {\bibfnamefont {Y.}~\bibnamefont {Ohnuma}}, \
  and\ \bibinfo {author} {\bibfnamefont {M.}~\bibnamefont {Matsuo}},\ }\href
  {\doibase 10.1103/PhysRevB.98.094430} {\bibfield  {journal} {\bibinfo
  {journal} {Phys. Rev. B}\ }\textbf {\bibinfo {volume} {98}},\ \bibinfo
  {pages} {094430} (\bibinfo {year} {2018})}\BibitemShut {NoStop}%
\bibitem [{\citenamefont {Smerzi}\ \emph {et~al.}(1997)\citenamefont {Smerzi},
  \citenamefont {Fantoni}, \citenamefont {Giovanazzi},\ and\ \citenamefont
  {Shenoy}}]{PhysRevLett.79.4950}%
  \BibitemOpen
  \bibfield  {author} {\bibinfo {author} {\bibfnamefont {A.}~\bibnamefont
  {Smerzi}}, \bibinfo {author} {\bibfnamefont {S.}~\bibnamefont {Fantoni}},
  \bibinfo {author} {\bibfnamefont {S.}~\bibnamefont {Giovanazzi}}, \ and\
  \bibinfo {author} {\bibfnamefont {S.~R.}\ \bibnamefont {Shenoy}},\ }\href
  {\doibase 10.1103/PhysRevLett.79.4950} {\bibfield  {journal} {\bibinfo
  {journal} {Phys. Rev. Lett.}\ }\textbf {\bibinfo {volume} {79}},\ \bibinfo
  {pages} {4950} (\bibinfo {year} {1997})}\BibitemShut {NoStop}%
\bibitem [{\citenamefont {Burchianti}\ \emph {et~al.}(2018)\citenamefont
  {Burchianti}, \citenamefont {Scazza}, \citenamefont {Amico}, \citenamefont
  {Valtolina}, \citenamefont {Seman}, \citenamefont {Fort}, \citenamefont
  {Zaccanti}, \citenamefont {Inguscio},\ and\ \citenamefont
  {Roati}}]{PhysRevLett.120.025302}%
  \BibitemOpen
  \bibfield  {author} {\bibinfo {author} {\bibfnamefont {A.}~\bibnamefont
  {Burchianti}}, \bibinfo {author} {\bibfnamefont {F.}~\bibnamefont {Scazza}},
  \bibinfo {author} {\bibfnamefont {A.}~\bibnamefont {Amico}}, \bibinfo
  {author} {\bibfnamefont {G.}~\bibnamefont {Valtolina}}, \bibinfo {author}
  {\bibfnamefont {J.~A.}\ \bibnamefont {Seman}}, \bibinfo {author}
  {\bibfnamefont {C.}~\bibnamefont {Fort}}, \bibinfo {author} {\bibfnamefont
  {M.}~\bibnamefont {Zaccanti}}, \bibinfo {author} {\bibfnamefont
  {M.}~\bibnamefont {Inguscio}}, \ and\ \bibinfo {author} {\bibfnamefont
  {G.}~\bibnamefont {Roati}},\ }\href {\doibase 10.1103/PhysRevLett.120.025302}
  {\bibfield  {journal} {\bibinfo  {journal} {Phys. Rev. Lett.}\ }\textbf
  {\bibinfo {volume} {120}},\ \bibinfo {pages} {025302} (\bibinfo {year}
  {2018})}\BibitemShut {NoStop}%
\bibitem [{\citenamefont {Gauthier}\ \emph {et~al.}(2019)\citenamefont
  {Gauthier}, \citenamefont {Szigeti}, \citenamefont {Reeves}, \citenamefont
  {Baker}, \citenamefont {Bell}, \citenamefont {Rubinsztein-Dunlop},
  \citenamefont {Davis},\ and\ \citenamefont {Neely}}]{PhysRevLett.123.260402}%
  \BibitemOpen
  \bibfield  {author} {\bibinfo {author} {\bibfnamefont {G.}~\bibnamefont
  {Gauthier}}, \bibinfo {author} {\bibfnamefont {S.~S.}\ \bibnamefont
  {Szigeti}}, \bibinfo {author} {\bibfnamefont {M.~T.}\ \bibnamefont {Reeves}},
  \bibinfo {author} {\bibfnamefont {M.}~\bibnamefont {Baker}}, \bibinfo
  {author} {\bibfnamefont {T.~A.}\ \bibnamefont {Bell}}, \bibinfo {author}
  {\bibfnamefont {H.}~\bibnamefont {Rubinsztein-Dunlop}}, \bibinfo {author}
  {\bibfnamefont {M.~J.}\ \bibnamefont {Davis}}, \ and\ \bibinfo {author}
  {\bibfnamefont {T.~W.}\ \bibnamefont {Neely}},\ }\href {\doibase
  10.1103/PhysRevLett.123.260402} {\bibfield  {journal} {\bibinfo  {journal}
  {Phys. Rev. Lett.}\ }\textbf {\bibinfo {volume} {123}},\ \bibinfo {pages}
  {260402} (\bibinfo {year} {2019})}\BibitemShut {NoStop}%
\bibitem [{Note3()}]{Note3}%
  \BibitemOpen
  \bibinfo {note} {In this case, the transition between Josephson oscillation
  and self-trapping regimes occurs when the ratio of the condensation energy to
  the Josephson coupling energy becomes $10-10^2$.}\BibitemShut {Stop}%
\bibitem [{\citenamefont {Grenier}\ \emph {et~al.}()\citenamefont {Grenier},
  \citenamefont {Kollath},\ and\ \citenamefont {Georges}}]{Grenier:aa}%
  \BibitemOpen
  \bibfield  {author} {\bibinfo {author} {\bibfnamefont {C.}~\bibnamefont
  {Grenier}}, \bibinfo {author} {\bibfnamefont {C.}~\bibnamefont {Kollath}}, \
  and\ \bibinfo {author} {\bibfnamefont {A.}~\bibnamefont {Georges}},\ }\href
  {\doibase http://dx.doi.org/10.1016/j.crhy.2016.08.013} {\bibinfo  {journal}
  {Comptes Rendus Physique}\ ,\ }\BibitemShut {NoStop}%
\bibitem [{\citenamefont {Mukherjee}\ \emph {et~al.}(2017)\citenamefont
  {Mukherjee}, \citenamefont {Yan}, \citenamefont {Patel}, \citenamefont
  {Hadzibabic}, \citenamefont {Yefsah}, \citenamefont {Struck},\ and\
  \citenamefont {Zwierlein}}]{Mukherjee:2017aa}%
  \BibitemOpen
\bibfield  {journal} {  }\bibfield  {author} {\bibinfo {author} {\bibfnamefont
  {B.}~\bibnamefont {Mukherjee}}, \bibinfo {author} {\bibfnamefont
  {Z.}~\bibnamefont {Yan}}, \bibinfo {author} {\bibfnamefont {P.~B.}\
  \bibnamefont {Patel}}, \bibinfo {author} {\bibfnamefont {Z.}~\bibnamefont
  {Hadzibabic}}, \bibinfo {author} {\bibfnamefont {T.}~\bibnamefont {Yefsah}},
  \bibinfo {author} {\bibfnamefont {J.}~\bibnamefont {Struck}}, \ and\ \bibinfo
  {author} {\bibfnamefont {M.~W.}\ \bibnamefont {Zwierlein}},\ }\href {\doibase
  10.1103/PhysRevLett.118.123401} {\bibfield  {journal} {\bibinfo  {journal}
  {Phys. Rev. Lett.}\ }\textbf {\bibinfo {volume} {118}},\ \bibinfo {pages}
  {123401} (\bibinfo {year} {2017})}\BibitemShut {NoStop}%
\bibitem [{\citenamefont {Luick}\ \emph {et~al.}(2019)\citenamefont {Luick},
  \citenamefont {Sobirey}, \citenamefont {Bohlen}, \citenamefont {Singh},
  \citenamefont {Mathey}, \citenamefont {Lompe},\ and\ \citenamefont
  {Moritz}}]{Luick:2019aa}%
  \BibitemOpen
  \bibfield  {author} {\bibinfo {author} {\bibfnamefont {N.}~\bibnamefont
  {Luick}}, \bibinfo {author} {\bibfnamefont {L.}~\bibnamefont {Sobirey}},
  \bibinfo {author} {\bibfnamefont {M.}~\bibnamefont {Bohlen}}, \bibinfo
  {author} {\bibfnamefont {V.~P.}\ \bibnamefont {Singh}}, \bibinfo {author}
  {\bibfnamefont {L.}~\bibnamefont {Mathey}}, \bibinfo {author} {\bibfnamefont
  {T.}~\bibnamefont {Lompe}}, \ and\ \bibinfo {author} {\bibfnamefont
  {H.}~\bibnamefont {Moritz}},\ }\href@noop {} {\bibfield  {journal} {\bibinfo
  {journal} {arXiv preprint arXiv:1908.09776}\ } (\bibinfo {year}
  {2019})}\BibitemShut {NoStop}%
\bibitem [{\citenamefont {Barone}\ and\ \citenamefont
  {Paterno}(1982)}]{barone1982}%
  \BibitemOpen
  \bibfield  {author} {\bibinfo {author} {\bibfnamefont {A.}~\bibnamefont
  {Barone}}\ and\ \bibinfo {author} {\bibfnamefont {G.}~\bibnamefont
  {Paterno}},\ }\href@noop {} {\emph {\bibinfo {title} {Physics and
  applications of the Josephson effect}}}\ (\bibinfo  {publisher} {Wiley},\
  \bibinfo {year} {1982})\BibitemShut {NoStop}%
\bibitem [{\citenamefont {Tinkham}(2004)}]{tinkham2004}%
  \BibitemOpen
  \bibfield  {author} {\bibinfo {author} {\bibfnamefont {M.}~\bibnamefont
  {Tinkham}},\ }\href@noop {} {\emph {\bibinfo {title} {Introduction to
  superconductivity}}}\ (\bibinfo  {publisher} {Courier Corporation},\ \bibinfo
  {year} {2004})\BibitemShut {NoStop}%
\bibitem [{\citenamefont {Karpiuk}\ \emph {et~al.}(2012)\citenamefont
  {Karpiuk}, \citenamefont {Gr\'emaud}, \citenamefont {Miniatura},\ and\
  \citenamefont {Gajda}}]{Karpiuk:2012aa}%
  \BibitemOpen
  \bibfield  {author} {\bibinfo {author} {\bibfnamefont {T.}~\bibnamefont
  {Karpiuk}}, \bibinfo {author} {\bibfnamefont {B.}~\bibnamefont {Gr\'emaud}},
  \bibinfo {author} {\bibfnamefont {C.}~\bibnamefont {Miniatura}}, \ and\
  \bibinfo {author} {\bibfnamefont {M.}~\bibnamefont {Gajda}},\ }\href
  {\doibase 10.1103/PhysRevA.86.033619} {\bibfield  {journal} {\bibinfo
  {journal} {Phys. Rev. A}\ }\textbf {\bibinfo {volume} {86}},\ \bibinfo
  {pages} {033619} (\bibinfo {year} {2012})}\BibitemShut {NoStop}%
\bibitem [{\citenamefont {Uchino}(2020)}]{uchino2020role}%
  \BibitemOpen
  \bibfield  {author} {\bibinfo {author} {\bibfnamefont {S.}~\bibnamefont
  {Uchino}},\ }\href@noop {} {\  (\bibinfo {year} {2020})},\ \Eprint
  {http://arxiv.org/abs/2003.08672} {arXiv:2003.08672 [cond-mat.quant-gas]}
  \BibitemShut {NoStop}%
\bibitem [{\citenamefont {Fornieri}\ and\ \citenamefont
  {Giazotto}(2017)}]{Fornieri:2017aa}%
  \BibitemOpen
  \bibfield  {author} {\bibinfo {author} {\bibfnamefont {A.}~\bibnamefont
  {Fornieri}}\ and\ \bibinfo {author} {\bibfnamefont {F.}~\bibnamefont
  {Giazotto}},\ }\href {http://dx.doi.org/10.1038/nnano.2017.204} {\bibfield
  {journal} {\bibinfo  {journal} {Nature Nanotechnology}\ }\textbf {\bibinfo
  {volume} {12}},\ \bibinfo {pages} {944 EP } (\bibinfo {year}
  {2017})}\BibitemShut {NoStop}%
\bibitem [{\citenamefont {Uchino}\ \emph {et~al.}(2018)\citenamefont {Uchino},
  \citenamefont {Ueda},\ and\ \citenamefont {Brantut}}]{PhysRevA.98.063619}%
  \BibitemOpen
  \bibfield  {author} {\bibinfo {author} {\bibfnamefont {S.}~\bibnamefont
  {Uchino}}, \bibinfo {author} {\bibfnamefont {M.}~\bibnamefont {Ueda}}, \ and\
  \bibinfo {author} {\bibfnamefont {J.-P.}\ \bibnamefont {Brantut}},\ }\href
  {\doibase 10.1103/PhysRevA.98.063619} {\bibfield  {journal} {\bibinfo
  {journal} {Phys. Rev. A}\ }\textbf {\bibinfo {volume} {98}},\ \bibinfo
  {pages} {063619} (\bibinfo {year} {2018})}\BibitemShut {NoStop}%
\bibitem [{Note4()}]{Note4}%
  \BibitemOpen
  \bibinfo {note} {We note $\DOTSI \intop \ilimits@ _{-\infty }^{\infty
  }\protect \frac {d\omega }{2\pi }\protect \frac {\omega ^3}{\protect \qopname
  \relax o{sinh}^2(\omega /(2T))}=0$}\BibitemShut {NoStop}%
\end{thebibliography}
%

\end{document}